\documentclass[a4paper,11pt]{article}
\pdfoutput=1 

\usepackage{jheppub}
\usepackage{float}
\usepackage{ulem}

\usepackage{mathrsfs}
\usepackage{setspace}
\usepackage{amsfonts}
\usepackage{amssymb}
\usepackage{amsmath}
\usepackage{esint}                   
\usepackage{epsfig}
\usepackage{latexsym}
\usepackage{color}
\usepackage{graphicx}
\usepackage{hyperref}
\usepackage{nicefrac}
\usepackage[latin1]{inputenc}
\usepackage{pstricks}
\usepackage{slashed}
\usepackage{multirow}
\usepackage{mathtools}
\usepackage[ddmmyyyy,hhmmss]{datetime}

\allowdisplaybreaks

\def\IR{\relax{\rm I\kern-.18em R}}

\begin{document}

\title{Holographic Krylov complexity in the Coulomb branch of ${\cal N}=4$ SYM}

\author[]{Dimitrios Zoakos}

\affiliation[]{Department of Physics, University of Patras, 26504 Patras, Greece.}

\emailAdd{dzoakos@upatras.gr}

\abstract{We study holographic Krylov complexity in the Coulomb branch of ${\cal N}=4$ SYM.
Adopting the proposal that the time derivative of the Krylov complexity is dual to the proper 
radial momentum of a massive particle, we investigate two probe geodesics within this geometry.
For one of the radial trajectories we obtain exact analytic results, 
even when additional motion in the internal space is included.
In cases where the geodesic avoids the interior curvature singularity, the Krylov complexity 
exhibits oscillatory behavior, with a frequency governed by the Coulomb scale 
and an amplitude determined by the UV cutoff, the Coulomb scale, and the angular momentum. 
This oscillatory pattern is lost, when the radial trajectory is approaching the singularity.
Finally, we compare our holographic results with field-theoretic calculations, finding qualitative agreement.
}

\maketitle

\flushbottom


\section{Introduction}

A significant recent development in the study of quantum many-body systems is the use of Krylov 
complexity to diagnose operator growth. 
In a QFT Krylov complexity characterizes the expansion of a time-evolved quantum state or operator 
within the Krylov subspace, which is generated by repeatedly applying the Hamiltonian or a 
Liouvillian on an initial operator or state \cite{Rabinovici:2025otw}. 
Recent research has significantly advanced the geometric understanding of Krylov complexity. 
While \cite{Caputa:2021sib} identifies its connection to geodesic volume and information geometry, 
\cite{Fan:2023ohh}  extends these concepts to a wider family of operator diagnostics. 
Furthermore, the momentum-Krylov correspondence proposed in \cite{Fan:2024iop}, establishes a 
link between the boundary growth rate of complexity and the radial momentum of in-falling particles in AdS black holes, 
a relation further scrutinized and refined in \cite{He:2024pox}. 
Together, these developments frame Krylov complexity as a robust geometric probe of chaotic evolution, 
paving the way for a deeper holographic interpretation.
Recently in \cite{Caputa:2024sux} a concrete formulation of this correspondence was proposed in the case of AdS backgrounds, 
where the rate of change of complexity is directly proportional to the proper momentum of a massive particle 
moving along the radial coordinate $\bar \rho$, namely
\begin{equation}
{\dot C}(t)= - \frac{P_{\bar {\rho}}}{\epsilon} \quad \text{where $\epsilon$ is a UV regulator.}
\end{equation}
This framework provides an appealing geometric interpretation of operator spreading, motivating further investigation into its applicability to holography, with a focus on strongly coupled theories.
In \cite{Fatemiabhari:2025cyy}, the concepts of proper momentum and proper coordinates were generalized to $AdS_5$
(sliced by $AdS_3$) to evaluate the evolution of holographic Krylov complexity at ${\cal N}=4$ SYM. 
This framework was further extended in \cite{Fatemiabhari:2025usn,Fatemiabhari:2026goj} 
to encompass confining gauge theories within top-down 
string duals.\footnote{For extensions to conformal quiver gauge theories and Yang-Baxter deformed supergravity backgrounds, see 
\cite{Fatemiabhari:2025poq,Fatemiabhari:2026rob} and \cite{Roychowdhury:2026eds}, respectively.}

The main scope of this paper is to use the prescription of \cite{Caputa:2024sux} to study the 
Krylov complexity in the gravity dual of the Coulomb branch 
of ${\cal N}=4$ SYM. The theory enters the Coulomb branch as the $SO(6)$ scalar fields acquire 
non-zero vacuum expectation values.  On the gravity side this correspond to multicenter distributions of 
branes and on the field theory side the boundary CFT is deformed by the VEV an operator with 
dimension-two that transforms in the $20'$ representation of $SO(6)$ and triggers a flow along the Coulomb branch \cite{Freedman:1999gp,Freedman:1999gk}. 
These flows are typically characterized by a singularity deep in the bulk and one of the objectives 
of this work is to probe this singularity through the lens of Krylov complexity. 
On the QFT side, the presence of the singularity suggests that the holographic dual fails to account for 
light fields arising at the Coulomb VEV scale, marking the limit of validity of the effective description.

To investigate Krylov complexity, we analyze the geodesic of a massive particle falling radially in the given background and calculate its proper momentum. We associate this geometric quantity 
with the time derivative of the Krylov complexity in the dual QFT. 
Furthermore, we examine the distinct behaviors of the proper momentum across different geodesics and 
extend our study to cases where the massive particle possesses angular momentum  via rotation in 
the internal space. The complexity exhibits an oscillatory behavior, where the frequency and amplitude 
are determined by the system's parameters--specifically, the UV cutoff, 
the Coulomb scale and the angular momentum.
A different behavior for the complexity arises, when the geodesic is approaching the singularity.

The plan of the paper is as follows. 
In section \ref{sec:typeIIB_background} we present the gravity solution, 
which serves as the background for our subsequent computation of the Krylov complexity. 
There are two possible orientations for the geodesic, 
and these are analyzed in the two subsections of 
section \ref{sec:Geodesic_motion}. 
In subsection \ref{subsec:Trajectory_theta_pi_over_2}  the geodesic is sitting at
$\theta=\pi/2$ and in subsection \ref{subsec:Trajectory_theta_0} at  $\theta=0$. 
In both cases the analysis 
of the basic configuration is extended to include particle motion in the internal space. 
The introduction of angular momentum -- a feature arising from the use of a top-down string dual -- 
leads to variations and we analyze the qualitative changes it produces.\footnote{This is analogous to the 
symmetry resolved Krylov complexity \cite{Caputa:2025mii,Caputa:2025ozd}.}
The proper momentum and the Krylov complexity are then detailed in section \ref{sec:PM_CX}, specifically in 
subsections \ref{subsec:PM_CX_theta_pi_over_2} and \ref{subsec:PM_CX_theta_0}, where we compare the 
differences and similarities between the two orientations. 
In subsection \ref{subsec:FTcomparison} we compare the gravity 
computation with a field theory calculation. 
Finally, section \ref{sec:conclusions} provides our conclusions and outlines potential future research directions.


\section{The type-IIB background}
\label{sec:typeIIB_background}

The starting point of the analysis will be the multicenter D3-brane solution of type-IIB
supergravity, originally derived in \cite{Kraus:1998hv,Russo:1998mm}. Its metric reads
\begin{eqnarray} \label{10d_metric}
ds^2& =&  \frac{\zeta}{z^2}\left[-dt^2+ d{\vec x}^2 +\frac{dz^2}{\lambda^6}\right]  + \zeta \, d\theta^2
+\frac{1}{\zeta}\Bigg[\cos^2\theta d\psi^2 + \\
& + & \cos^2\theta\sin^2\psi \, d\phi_1^2 +
\cos^2\theta\cos^2\psi \, d\phi_2^2 +\lambda^6 \sin^2\theta \, d\phi_3^2\Bigg].\nonumber
\end{eqnarray}
where the functions $\zeta$ and $\lambda$ are defined as follows 
\begin{equation}
\zeta(z , \theta)^2 = 1 - \ell^2 \, z^2 \cos^2\theta \quad \& \quad  \lambda(z)^6 = 1 - {\ell^2} {z^2}
\quad {\rm with} \quad 0<z<1/\ell \, . 
\end{equation}
The  geometry terminates at $z=1/\ell $, where a curvature singularity occurs. 

To investigate the existence of singularities, one must examine the behavior of the curvature 
invariants close to the end of space, i.e. $z=1/\ell $.  As discussed in \cite{Chatzis:2025dnu, Chatzis:2025hek} 
(and previously explored in \cite{Brandhuber:1999jr, Hernandez:2005xd}), the invariants $R_{ab}R^{ab}$ and 
$R_{abcd}R^{abcd}$ exhibit a divergent behavior as $\theta \rightarrow 0$.  This divergence indicates 
that higher-derivative corrections become non-negligible, signaling the breakdown of the supergravity 
approximation in this regime.

In what follows, we explicitly perform the holographic calculation of the Krylov complexity in a conformal theory 
deformed by the VEV of a dimension-two operator. 
One of the objectives is to 
examine the impact of the bulk singularity as the dual geodesic probes the deep interior.
It is for this reason that we consider a geodesic that is sitting at $\theta = 0$. 
This study departs from previous works, such as \cite{Caputa:2024sux}, in two key aspects: 
it employs a top-down construction and allows the geodesic to probe the $S^5$ degrees of freedom. 
While we recover the $SL(2)$ symmetry inside $AdS_5$  in the UV regime--consistent with \cite{Caputa:2024sux}--
our results deviate significantly in the IR limit ($z\sim1/\ell $). 
To conduct this analysis, we examine a probe particle falling within the Coulomb branch background given by \eqref{10d_metric}.


\section{Geodesic motion}
\label{sec:Geodesic_motion}

We now consider a particle in the background \eqref{10d_metric}, with a geodesic that also probes the 
$S^{5}$ interior. Accordingly, the radial trajectory is parameterized by two time-dependent coordinates:
the holographic radial coordinate and an angle within the $S^{5}$. 
Two distinct choices for the angular coordinate exist: the first involves setting $\theta =\pi /2$, 
while the second consists of the configuration where $\theta =0$.\footnote{
Notice that a similar possibility for two different string configurations, 
within the Coulomb branch, appears in the 
Wilson loop computations \cite{Brandhuber:1999jr, Chatzis:2025dnu,Chatzis:2025hek}.} 
In what follows, we analyze the resulting geodesics for both cases.


\subsection{Trajectory with $\theta=\pi/2$}
\label{subsec:Trajectory_theta_pi_over_2}

We consider the following ansatz for the geodesics
\begin{equation}
t=\tau\, ,  \quad \theta = \frac{\pi}{2}  \, , \quad z = z(\tau) \quad \& \quad \phi_3 = \phi(\tau) 
\end{equation}
where all the other coordinates are held fixed.
Notice that this ansatz is consistent and satisfies the equations of motion, provided that $\theta =\pi /2$.
The induced metric for the particle is 
\begin{equation} \label{induced_metric}
ds_{ind}^2 =- \frac{\zeta \, d\tau^2}{z^2} \Bigg[1- \frac{z^2 \, \lambda^6 \, \dot{\phi}_3^2}{\zeta^2}- \frac{\dot{z}^2}{\lambda^6}\Bigg]  \, .
\end{equation}
Evaluating $\zeta$ at $\theta =\pi /2$ yields $\zeta =1$. Consequently, this factor does not appear in the remaining derivations of this subsection.

The action for the point particle is
\begin{equation}
\mathcal{S}_P\sim \int d\tau \mathcal{L}_P \quad {\rm with} \quad 
\mathcal{L}_P=\frac{1}{z}\, \sqrt{1- z^2 \, \lambda^6 \, \dot{\phi}^2 - \frac{\dot{z}^2}{\lambda^6}}
\end{equation}
and the Lagrangian has the following two conserved quantities 
\begin{equation}
 L_\phi = \frac{z \, \lambda^6 \, \dot{\phi}}{\sqrt{1- z^2 \, \lambda^6 \, \dot{\phi}^2 - \frac{\dot{z}^2}{\lambda^6}}}
 \quad \& \quad  
 \mathcal{H}=\frac{1}{z \sqrt{1- z^2 \, \lambda^6 \, \dot{\phi}^2 - \frac{\dot{z}^2}{\lambda^6}}}
\end{equation}
namely the angular momentum, that is associated with translations along the $\phi_3$-cycle, and the 
Hamiltonian, that is associated with the time translation symmetry. Solving those relations with respect to 
$\dot{z}(t)$ and $\dot{\phi}(t)$ we obtain the following expressions
\begin{equation} \label{z_phi_dot_v1}
\dot{z}(t)= \pm \,\frac{\lambda^3}{\mathcal{H} \, z}\sqrt{\mathcal{H}^2 \, z^2- 1 -  \frac{L_\phi^2}
{\lambda^6} }  \quad \& \quad  
{\dot \phi} = \frac{L_\phi}{\mathcal{H} \, z^2 \, \lambda^6} \, . 
\end{equation}
To facilitate the integration with respect to $z$ we are rewriting $\dot{z}(t)$ in the following way
\begin{equation} \label{z_dot_v2}
\dot{z}(t)= \pm \,  \frac{\ell}{z} \sqrt{\left({z^2}-{z_{-}^2} \right) \left(z_{+}^2-z^2 \right)} 
\quad {\rm with} \quad z \in [ z_{-} , z_{+}]
\end{equation}
where $z_{-} $ and $z_{+} $ are defined as follows
\begin{equation} \label{def_zpm}
z_{\pm}^2 = \frac{1}{2\, \mathcal{H}^2 \ \ell^2 } \Bigg[\mathcal{H}^2 + \ell^2 \pm \sqrt{\left(\mathcal{H}^2 - \ell^2\right)^2 - 4 \, \mathcal{H}^2\, \ell^2 \, L_\phi^2 }\Bigg] \, . 
\end{equation}
These are the values of $z$ that the first derivative with respect to the time $t$ vanishes, 
and denote the positions of the turning points for the particle trajectory inside the bulk. Notice that in the limit 
of $L_\phi \rightarrow 0$ we have that $z_-\rightarrow \mathcal{H}^{-1}$ and 
$z_+\rightarrow \ell^{-1}$. This means that a geodesic that is rotating inside the $S^5$ will never probe 
the ``end of space", which is located at $\ell^{-1}$.

Integrating \eqref{z_dot_v2} we are able to compute analytically the relation between $t$ and $z$
\begin{equation} \label{integral_z}
 \int_0^t d\tau =  \frac{1}{\ell}  \int_{z_-}^z  
 \frac{z\, dz}{\sqrt{\left({z^2}- {z_{-}^2} \right) \left(z_{+}^2-{z^2} \right)}}
 \quad \Rightarrow \quad t= \frac{\pi }{ 2 \, \ell} - \frac{1}{\ell} \, \arctan \left[\frac{z_+^2 - z^2}{z^2 - z_-^2}\right] 
\end{equation}
and we have used the boundary condition that at $t=0$ the particle is at $z=z_-$. 
Notice that the time needed to travel from $z_-$ to $z_+$ is $\pi /(2 \, \ell)$. The relation between $t$ and $z$ 
in \eqref{integral_z} can be inverted, and after using the definitions for  $z_{-} $ and $z_{+} $ from 
\eqref{def_zpm}, we arrive to the following expression
\begin{equation} \label{z_function_t}
 z(t) = \frac{1}{\sqrt{2} \,\mathcal{H} \, \ell } \sqrt{\mathcal{H}^2 + \ell^2 -
 \cos (2 \, t \, \ell) \, \sqrt{\left(\mathcal{H}^2 - \ell^2\right)^2 - 4 \, \mathcal{H}^2\, \ell^2 \, L_\phi^2}} \, .
\end{equation}

\begin{figure}[ht] 
   \centering
   \includegraphics[width=7cm,height=7cm]{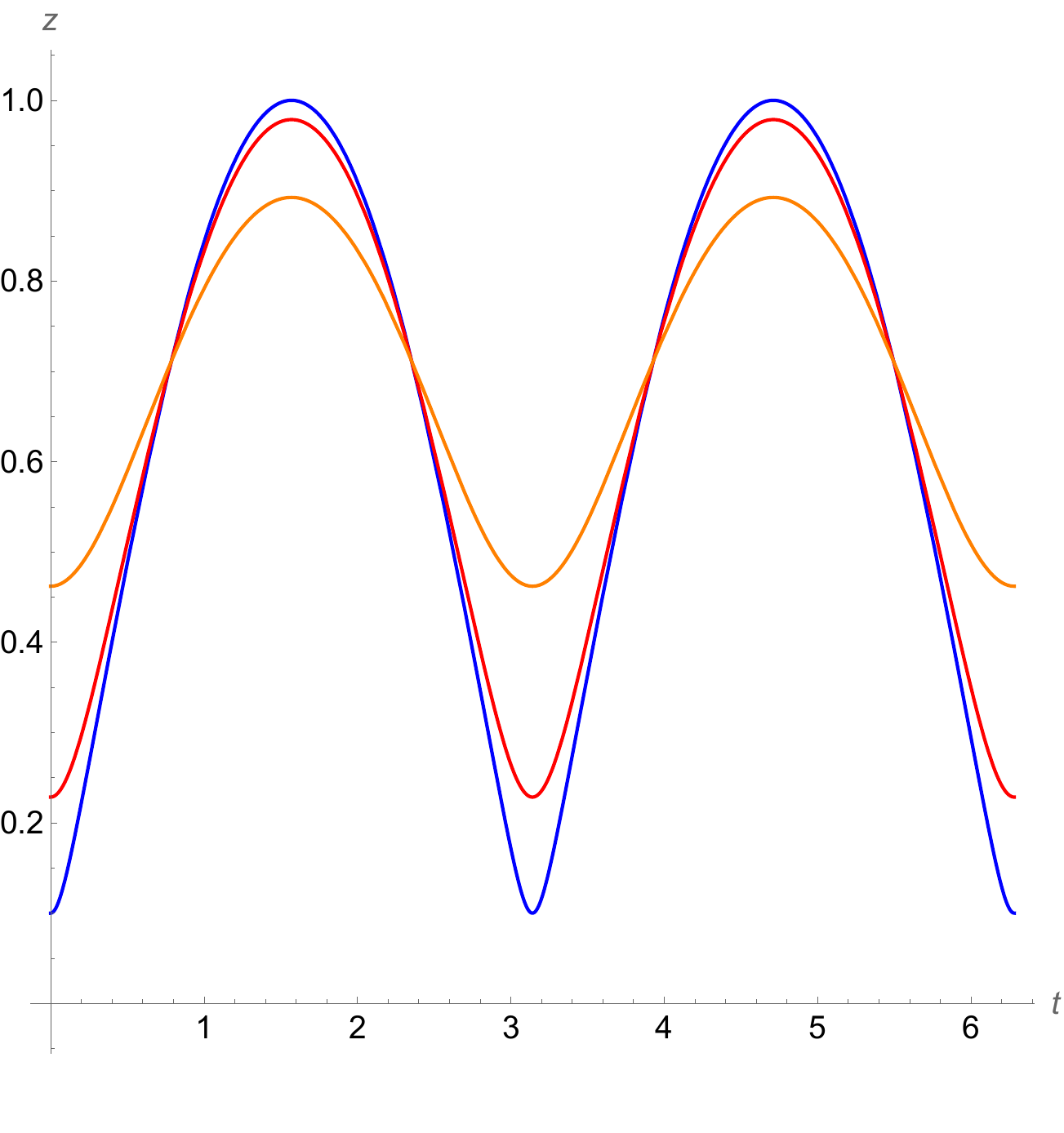}
    \includegraphics[width=7cm,height=7cm]{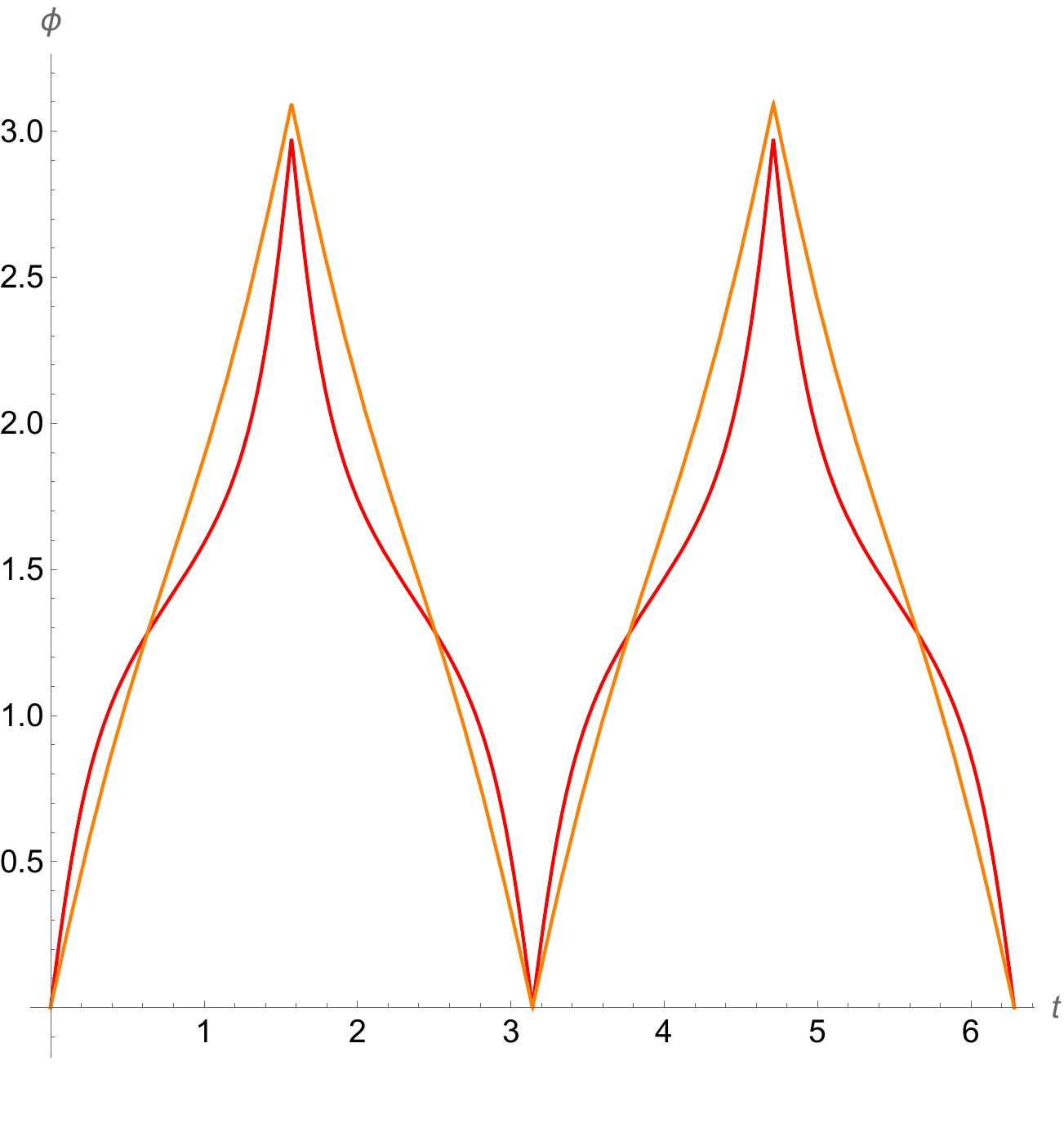}
     \caption{The left panel plots the function $z(t)$ for different values of $L_{\phi}$ 
     (blue for $L_{\phi}=0$, red for $L_{\phi}=2$, orange for $L_{\phi}=4$), 
     with  $\mathcal{H}=10$ and $\ell=1$. 
     The right panel shows the corresponding $\phi(t)$ evolution using the same parameter values.}
   \label{z_angle_theta_pi/2}
\end{figure}

In the limit of $L_\phi \rightarrow 0$ and for $\mathcal{H} > \ell$ the above expression simplifies to
\begin{equation} \label{z_function_t_no_rotation}
 z(t) = \sqrt{\frac{ \cos^2 (t \, \ell)}{\mathcal{H}^2} + \frac{ \sin^2 (t \, \ell)}{\ell^2} } \, .
\end{equation}
Notice that for $\mathcal{H} = \ell$, the UV cutoff and the end of space coincide;  consequently 
$z$ becomes constant (i.e. $z=1/\mathcal{H} $) and the travel time for the massive particle vanishes. 
Conversely, in the limit $\ell\rightarrow 0$, the travel time diverges, 
consistent with the time required for a massive particle to reach the horizon from the AdS boundary. 
Furthermore, the limit  $\ell\rightarrow 0$ of equation \eqref{z_function_t_no_rotation}
coincides with the AdS result obtained in \cite{Caputa:2024sux}, namely
\begin{equation}
z(t) = \sqrt{\big.{\epsilon}^2 + t^2} \quad {\rm with } \quad  \mathcal{H} = \frac{1}{\epsilon}\, .
\end{equation}
In the left panel of figure  \ref{z_angle_theta_pi/2} we plot the particle trajectory as a function of time for three 
values of the angular momentum. Some important comments are in order. 
The parameter $\ell $, which characterizes the background deformation, introduces a finite 
radial cut-off for the particle trajectory. 
In the non-rotating limit, the maximum radial extension is $z_{max}=\ell ^{-1}$. 
Consequently, increasing $\ell$ reduces the geodesic distance of the massive particle's trajectory in the bulk.
Similar behavior was reported in \cite{Roychowdhury:2026eds} and \cite{Fatemiabhari:2025usn}, 
arising from the Yang-Baxter deformation and the introduction of a parameter for confinement, respectively. 
The implications of this deformation for the growth of complexity will be discussed in the following section.
The introduction of rotation further deforms the particle trajectory. As illustrated in figure 
\ref{z_angle_theta_pi/2}, increasing $L_{\phi} $ further diminishes the geodesic distance of the particle's path, 
thereby narrowing the interval between the minimum and maximum values of $z$. 

In conclusion, when there is motion of the particle in the internal space the geodesic 
will never reach the point where the geometry terminates, i.e. $z=1/\ell$. 
It should be noted that even if the particle reaches the terminal point of the geometry, it is not probing the 
singularity.  The geodesic is restricted from approaching the singularity,  since it is sitting at $\theta = \pi/2$.
In all cases, whether considering the effects of $\ell$ alone or in combination with $L_{\phi }$, 
the radial profile exhibits a sinusoidal behavior bounded between a minimum and a maximum value, a feature also reported in \cite{Fatemiabhari:2025usn}. 

Combining the expressions for $\dot z$ and $\dot \phi$ from equation \eqref{z_phi_dot_v1}, we obtain
\begin{equation} \label{dphi_over_dz_theta_pi/2}
\frac{d \phi}{dz} = \frac{L_\phi}{\ell \, \mathcal{H}}
\frac{1}{z (1- \ell^2 \, z^2 )}
\frac{1}{\sqrt{\left({z^2}- {z_{-}^2} \right) \left(z_{+}^2-{z^2} \right)}}
\end{equation}
Performing the integration allows us to express $\phi$ as a function of $z$. 
The result of the integrations reads
\begin{equation}
\phi(z)=  \frac{L_\phi}{\sqrt{1+L_\phi^2}} \Bigg[\frac{\pi}{2} - \arctan \left[\frac{z_{-}}{z_{+}}  
\sqrt{ \frac{z_{+}^2- z^2}{z^2-z_{-}^2}}\right]\Bigg] + \frac{\pi}{2} -
\arctan \Bigg[\sqrt{\frac{z_{+}^2 - z^2}{z^2-z_{-}^2} \frac{1- \ell^2 \, z_{-}^2 }{1- \ell^2 \, z_{+}^2}}\Bigg]. 
\end{equation}
Subsequently, we substitute the expression for $z(t)$ from \eqref{z_function_t} into the angular evolution and 
in the right panel of figure \ref{z_angle_theta_pi/2} we plot the angular trajectory within the sphere as a function of time for two distinct values of $L_{\phi }$. The limiting values for \(\phi \) are given by
\begin{equation}
\phi\left(t=\frac{\pi}{2\, \ell}\right)  = 
\frac{\pi}{2}\left[1+\frac{L_\phi}{\sqrt{1+L_\phi^2}}\right] \quad \& \quad 
\phi\left(t=0\right) =0 \, .  
\end{equation}

\subsection{Trajectory with $\theta=0$}
\label{subsec:Trajectory_theta_0}

An alternative choice for the particle geodesic trajectory is the following 
\begin{equation}
t=\tau\, ,  \quad \theta = 0 \, , \quad z = z(\tau) \quad \& \quad \phi_1=\phi_2 = \phi(\tau)
\end{equation}
where all the other coordinates are held fixed. It should be emphasized that this ansatz is consistent, 
as the equations of motion for the particle are satisfied under the condition $\theta =0$.

\begin{figure}[ht] 
   \centering
   \includegraphics[width=7cm,height=7cm]{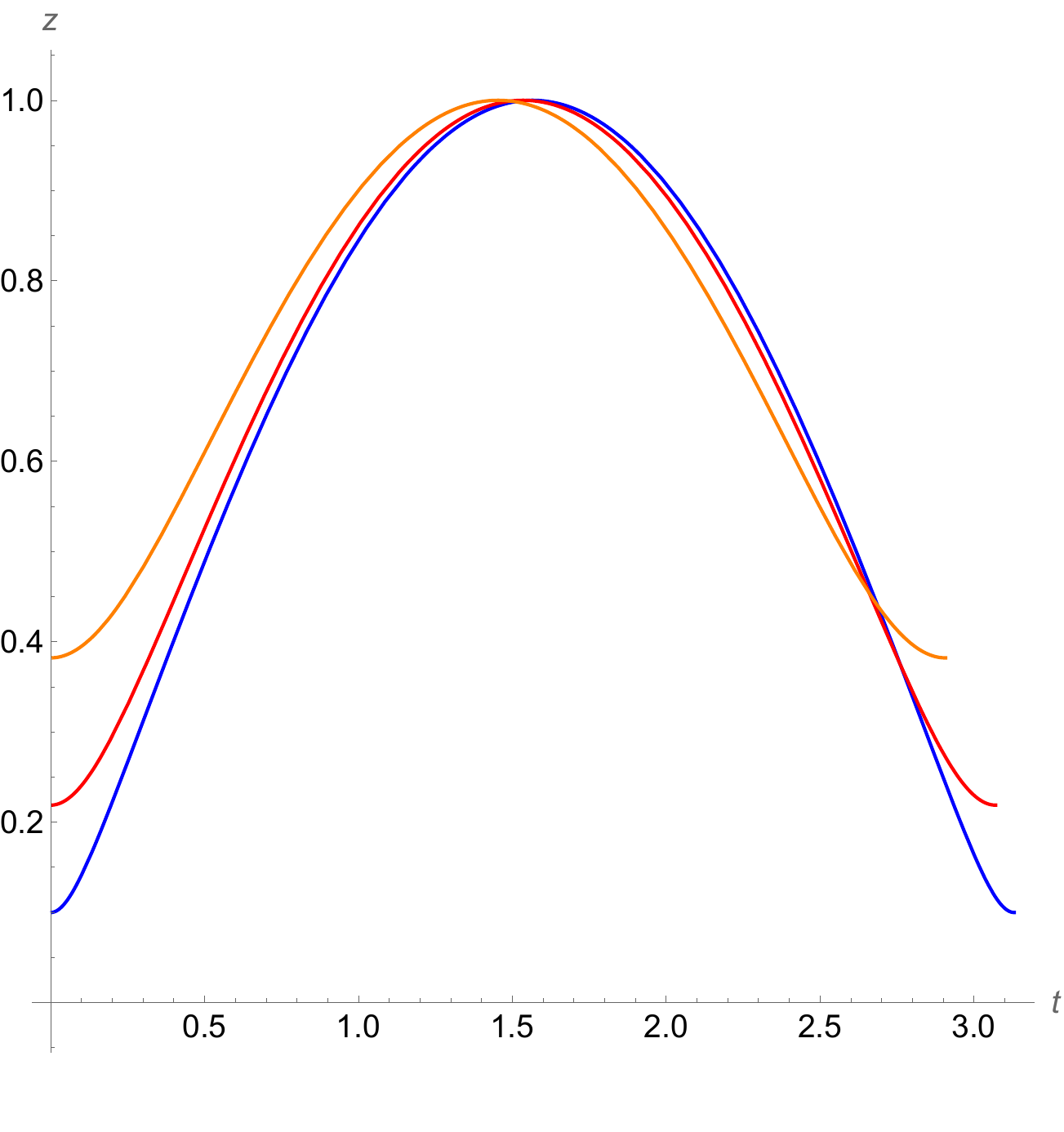}
    \includegraphics[width=7cm,height=7cm]{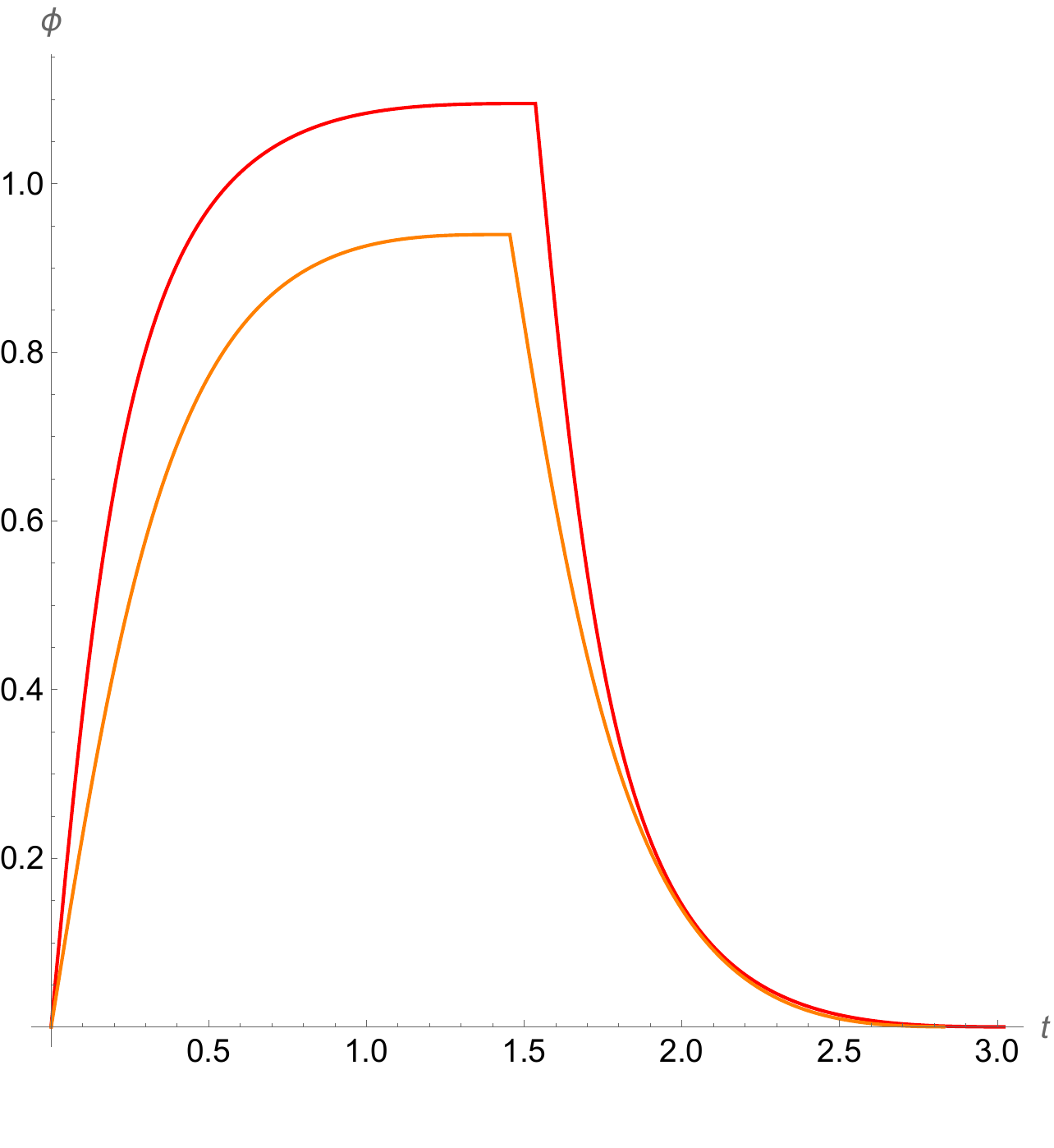}
     \caption{The left panel plots the function $z(t)$ for different values of $L_{\phi}$ 
     (blue for $L_{\phi}=0$, red for $L_{\phi}=2$, orange for $L_{\phi}=4$), 
     with  $\mathcal{H}=10$ and $\ell=1$. 
     The right panel shows the corresponding $\phi(t)$ evolution using the same parameter values.}
   \label{z_angle_theta_0}
\end{figure}

The action for the point particle is 
\begin{equation}
\mathcal{S}_P\sim \int d\tau \mathcal{L}_P \quad {\rm with} \quad 
\mathcal{L}_P=\frac{\sqrt{\zeta}}{z}\, 
\sqrt{1- \frac{z^2}{\zeta^2} \, \dot{\phi}^2- \frac{\dot{z}^2}{\lambda^6}}
\end{equation}
where the function $\zeta$, here and in the following expressions, is evaluated at $\theta =0$.
Notice that contrary to the $\theta =\pi/2$ case, its value is different from unity, and for this reason 
it will appear explicitly in the expressions of this subsection. 
The Lagrangian has two conserved quantities, namely the energy and the angular momentum, and it is possible 
to express $\dot{z}(t)$ and ${\dot \phi}(t) $ in terms of those conserved quantities, as follows
\begin{equation}  \label{z_phi_theta_0_dot_v1}
\dot{z}(t)= \pm \frac{\lambda^3}{\mathcal{H} \, z}
\sqrt{\mathcal{H}^2 \, z^2-\zeta (1+   L_\phi^2 \,\zeta)}
 \quad \& \quad  
{\dot \phi} = \frac{L_\phi \, \zeta^2}{\mathcal{H} \, z^2} \, . 
\end{equation}
Integrating the first of the equations above it is possible to obtain an analytic expression relating $t$ and $z$.
However, unlike the $\theta=\pi/2$ case, this relation cannot be inverted and is therefore omitted.
The integral expression reads
\begin{eqnarray} \label{integral_z_theta_0}
 \int_0^t d\tau &=&   \int_{z_-}^z  
 \frac{\mathcal{H}\, z\, dz}{\sqrt{\left(1- \ell^2 \, {z^2} \right) 
 \left[\mathcal{H}^2 \, z^2 - 
 \sqrt{1- \ell^2 \, {z^2} } - 
 L_\phi^2 \,\left(1- \ell^2 \, {z^2} \right) \right]}}
 \quad {\rm with} \quad
 \nonumber \\ 
  z_-& =&  \frac{1}{\left(\mathcal{H}^2 + \ell^2 \, L_\phi^2 \right)}
  \sqrt{\mathcal{H}^2\, L_\phi^2 - \ell^2\left(\frac{1}{2}- L_\phi^4\right) +
  \sqrt{\mathcal{H}^4+\frac{\ell^4}{4} + \mathcal{H}^2 \, \ell^2 \,L_\phi^2}}
\end{eqnarray}
and the maximum value for $z$ is $1/\ell$. 
In the left panel of figure  \ref{z_angle_theta_0} we plot the particle trajectory as a function of time for three 
values of the angular momentum. 
Comparing this figure with figure \ref{z_angle_theta_pi/2}, a major difference emerges for the particle 
trajectory between the two orientations. While in the $\theta=\pi/2$ case, 
the motion in the internal space prevents the particle from approaching the terminal point of the geometry, in the 
$\theta=0$ case rotation in the internal space is only affecting the initial value of the $z$ coordinate.
Given  the singularity that arises as $\theta \rightarrow 0$ and $z \rightarrow 1/\ell$, 
it is evident that such a geodesic will approach the singularity, regardless of the motion in the internal space. 
In the following section we examine the behavior of the Krylov complexity probe when the particle follows such a trajectory.  

When there is no motion in the internal space, it is possible to invert the relation between $z$ and $t$ and 
obtain analytic expressions of $z$ as a function of $t$. Those expressions are 
\begin{eqnarray} \label{z_approx_no_rotation_theta_0}
z(t) &=& \frac{1}{\ell} - \frac{\ell}{2}\left(t-t_e\right)^2 +\cdots 
\quad {\rm with } \quad t_e = \frac{1}{\ell} \, 
\arctan \left(\frac{2 \, \mathcal{H}^2}{\ell^2}\right)
\nonumber \\
z(t) &=&  z^{0}_- +\frac{1}{ 4\, z^{0}_-} \Big[2- \ell^2 \,(z^{0}_-)^2 \Big]^2 \, t^2 \cdots
\end{eqnarray}
where $ z^{0}_- $ is the starting point of the particle trajectory
\begin{equation} \label{z0initial_no_rotation_theta_0}
 z^{0}_- =  \frac{1}{\sqrt{2 }\, \mathcal{H}^2}
 \sqrt{\sqrt{4\, \mathcal{H}^4 + \ell^4}-\ell^2} 
\end{equation} 
and $t_e$ is the time that the particle needs in order to travel between the minimum and the maximum value 
of $z$ in the bulk. Another difference from the $\theta=\pi/2$ configuration arises here: 
While in the $\theta=\pi/2$ case the travel time for the particle between the endpoints of the particle trajectory is always $\pi/2 \ell$, 
in the $\theta=0$ case this time interval depends also on $\mathcal{H}$ and $L_{\phi}$. Specifically, when $L_{\phi}=0$, the
time is given by $t_e$. This dependence will be reflected in the Krylov complexity profiles presented in the following section.

Combining the expressions for $\dot z$ and $\dot \phi$ from equation \eqref{z_phi_theta_0_dot_v1}, we obtain
\begin{equation} \label{dotphi_over_dotz_theta_0}
\frac{d \phi}{dz} =
\frac{L_{\phi} \sqrt{1- \ell^2 \, {z^2} } }{z\, \sqrt{
\mathcal{H}^2 \, z^2 - \sqrt{1- \ell^2 \, {z^2} } - 
L_\phi^2 \,\left(1- \ell^2 \, {z^2} \right)}} \, . 
\end{equation}
After integrating to find $\phi$ as a function of $z$, we substitute the previously derived $z(t)$
relation to determine the angular trajectory on the sphere as a function of time. 
In the right panel of figure \ref{z_angle_theta_0} we plot the angular trajectory within the sphere as a function of time for two distinct values of $L_{\phi }$. Notice that as dictated by equation
\eqref{z_phi_theta_0_dot_v1}, 
$\dot \phi \rightarrow \frac{L_\phi \, \sqrt{1-\ell^2 ( z^{0}_- )^2 } }{\mathcal{H} \,( z^{0}_- )^2 } $ 
as $t\rightarrow 0$ and
$\dot \phi \rightarrow 0$ as $z\rightarrow 1/\ell$.

\section{The proper momentum and complexity}
\label{sec:PM_CX}

In accordance with \cite{Caputa:2024sux}, we introduce a new radial coordinate, denoted by $ \bar \rho$, 
which is conjectured to correspond to the Krylov basis in the dual field theory. 
This coordinate is defined as a \textit{proper radial distance} by imposing the constraints 
$\Delta t= \Delta {\vec x} =\Delta \theta =\Delta \psi =\Delta \phi_{i}=0$ (for appropriate indices $i$).
Specifically, we set $\Delta \phi _{1}=\Delta \phi _{2}=0$ for a trajectory at $\theta =\pi /2$, and 
$\Delta \phi _{3}=0$ for one at $\theta =0$. Then for both cases $\Delta s =\Delta {\bar \rho}$ or 
$d s^2 =d{\bar \rho}^2$.\footnote{In \cite{Caputa:2024sux}, the massive particle is falling along a
timelike geodesic in an asymptotically $AdS_3$ space (BTZ or global AdS) and the authors propose
the radial coordinate $\bar \rho$ to measure proper radial distance (either to the horizon of BTZ, or to the 
Poincare ``horizon" in the case of AdS).}
In the following, we analyze both configurations in detail.


\subsection{Trajectory with $\theta=\pi/2$}
\label{subsec:PM_CX_theta_pi_over_2}

For a radial trajectory at $\theta=\pi/2$, the geodesic distance in \eqref{10d_metric} simplifies to\footnote{Notice that $\phi$ is a function of $z$ (see equation \eqref{dphi_over_dz_theta_pi/2}).} (see also \cite{Fatemiabhari:2025cyy,Fatemiabhari:2026goj})
\begin{equation} \label{geo_distance}
ds^2 =   \frac{1}{z^2} \frac{dz^2}{\lambda^6}    +\lambda^6 \, d\phi^2 \equiv  d{\bar \rho}^2 
\end{equation}

In the absence of rotation the above relation becomes 
\begin{equation}
d{\bar \rho}^2 =   \frac{1}{z^2} \frac{dz^2}{\lambda^6} \quad \Rightarrow \quad 
d{\bar \rho} = -  \frac{1}{z} \frac{dz}{\sqrt{1 - \ell^2 \, z^2}} \quad \Rightarrow \quad 
z= \frac{1}{\ell \, \cosh {\bar \rho}} \, .
\end{equation}
Similarly to \cite{Fatemiabhari:2025usn} we have chosen a minus sign for $d{\bar \rho} $ to guarantee that $z$
is decreasing when $\bar \rho$ is increasing. The limit ${\bar \rho} \rightarrow \infty$ maps to the UV boundary where 
$z \sim e^{-{\bar \rho }}$ holds, aligning with the Krylov basis mapping introduced in \cite{Caputa:2024sux}. 
On the other hand, $\bar \rho \rightarrow 0$ corresponds to the IR scale $z_*=\frac{1}{\ell}$

\begin{figure}[ht] 
   \centering
   \includegraphics[width=7cm,height=7cm]{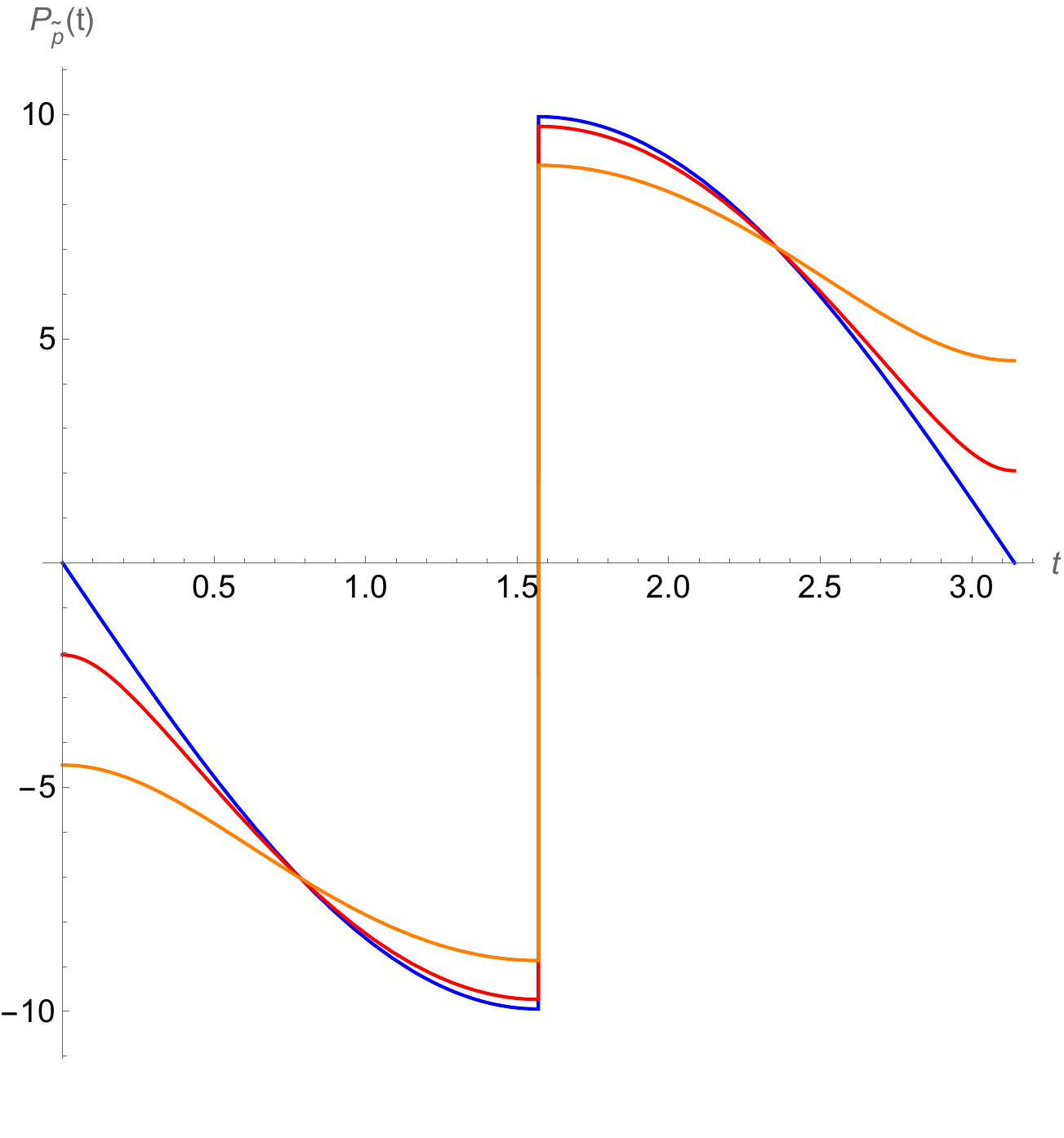}
    \includegraphics[width=7cm,height=7cm]{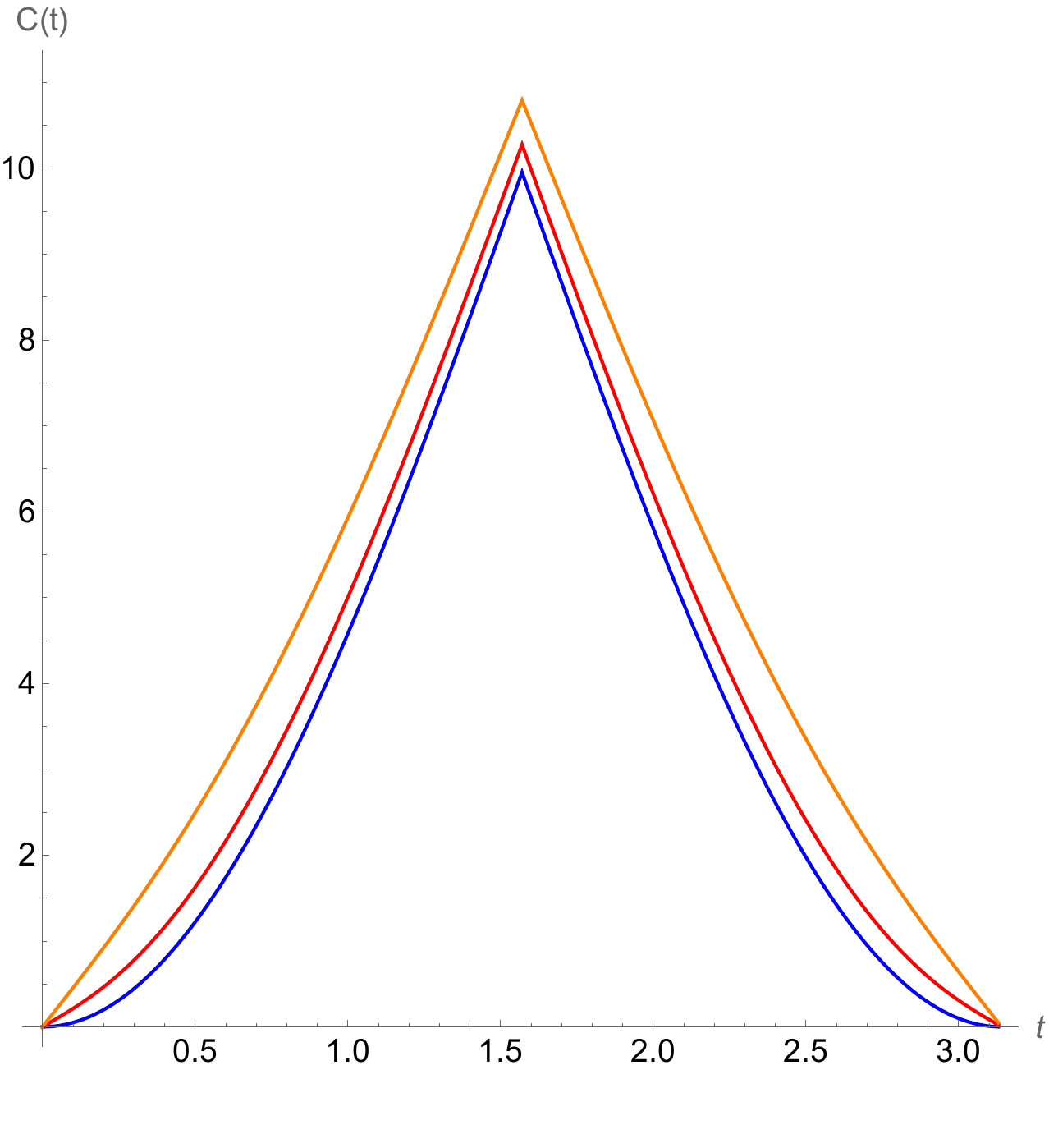}
     \caption{The left panel plots $P_{\bar \rho} $ and the right panel ${\cal C}(t)$ 
     for different values of $L_{\phi}$ (blue for $L_{\phi}=0$, red for $L_{\phi}=2$, 
     orange for $L_{\phi}=4$), with  $\mathcal{H}=10$ and $\ell=1$.}
   \label{PM_CX_theta_pi/2}
\end{figure}

From  \eqref{geo_distance} we have that 
\begin{equation}
\frac{d \bar \rho}{d z} = \sqrt{ \frac{1}{z^2} \frac{1}{\lambda^6}  +\lambda^6 \, \left(\frac{d\phi}{dz}\right)^2} 
\quad \& \quad 
\frac{d \bar \rho}{d \phi} = \sqrt{ \frac{1}{z^2} \frac{1}{\lambda^6}  \left(\frac{dz}{d\phi}\right)^2   +\lambda^6} 
\end{equation}
and the proper momentum is given by the following relation\footnote{The functions $z(t)$ and $\phi(t)$ are monotonic in the intervals $[0,\pi/2 \ell]$ and $[\pi/2 \ell, \pi/\ell]$.}
\begin{equation} \label{prop_mom_v1}
P_{\bar \rho} = P_z \, \frac{\partial {\dot z}}{ \partial {\dot {\bar \rho}}} +P_{\phi}   \, 
\frac{\partial {\dot \phi}}{ \partial {\dot {\bar \rho}}} \quad \& \quad  
\frac{dz}{d\phi} = \frac{\dot z}{{\dot \phi}} \, . 
\end{equation}
Substituting the expressions for the canonical momenta associated with the variables $z(t)$ and $\phi(t)$, namely
\begin{equation}
P_z = \frac{\partial \mathcal{L}_P}{\partial {\dot z}} = - \frac{\dot z}{ z^2 (1 - \ell^2 \, z^2) \mathcal{L}_P}  \quad \& \quad 
P_{\phi} = \frac{\partial \mathcal{L}_P}{\partial {\dot \phi}} = - \frac{ (1 - \ell^2 \, z^2)  \dot \phi}{\mathcal{L}_P} 
\end{equation}
into \eqref{prop_mom_v1}, we arrive to the following expression
\begin{equation} \label{prop_mom_v2}
P_{\bar \rho} = - \sqrt{\mathcal{H}^2 \, z^2 -1} \, . 
\end{equation}
Notice that the proper momentum does not depend explicitly on the angular momentum, 
however, an implicit dependence remains through the functional form of $z(t)$.\footnote{
This observation is in agreement with the similar calculation of the proper momentum, 
under the presence of angular momentum, in \cite{Fatemiabhari:2026goj}.}
In the left panel of figure \ref{PM_CX_theta_pi/2} we have plotted the proper momentum as a function 
of time for different values of the angular momentum $L_{\phi}$. This plot demonstrates an increase at early 
times and a slight decrease at late times (as $t \rightarrow \pi/2 \ell$) as 
the angular momentum increases, leading to a subsequent increase in complexity.\footnote{This finding is 
in agreement with the results in \cite{Fatemiabhari:2026goj}, 
noting that in their case the angular momentum is for a 
motion within AdS rather than the internal space.} 

\begin{figure}[ht] 
   \centering
   \includegraphics[width=7cm,height=7cm]{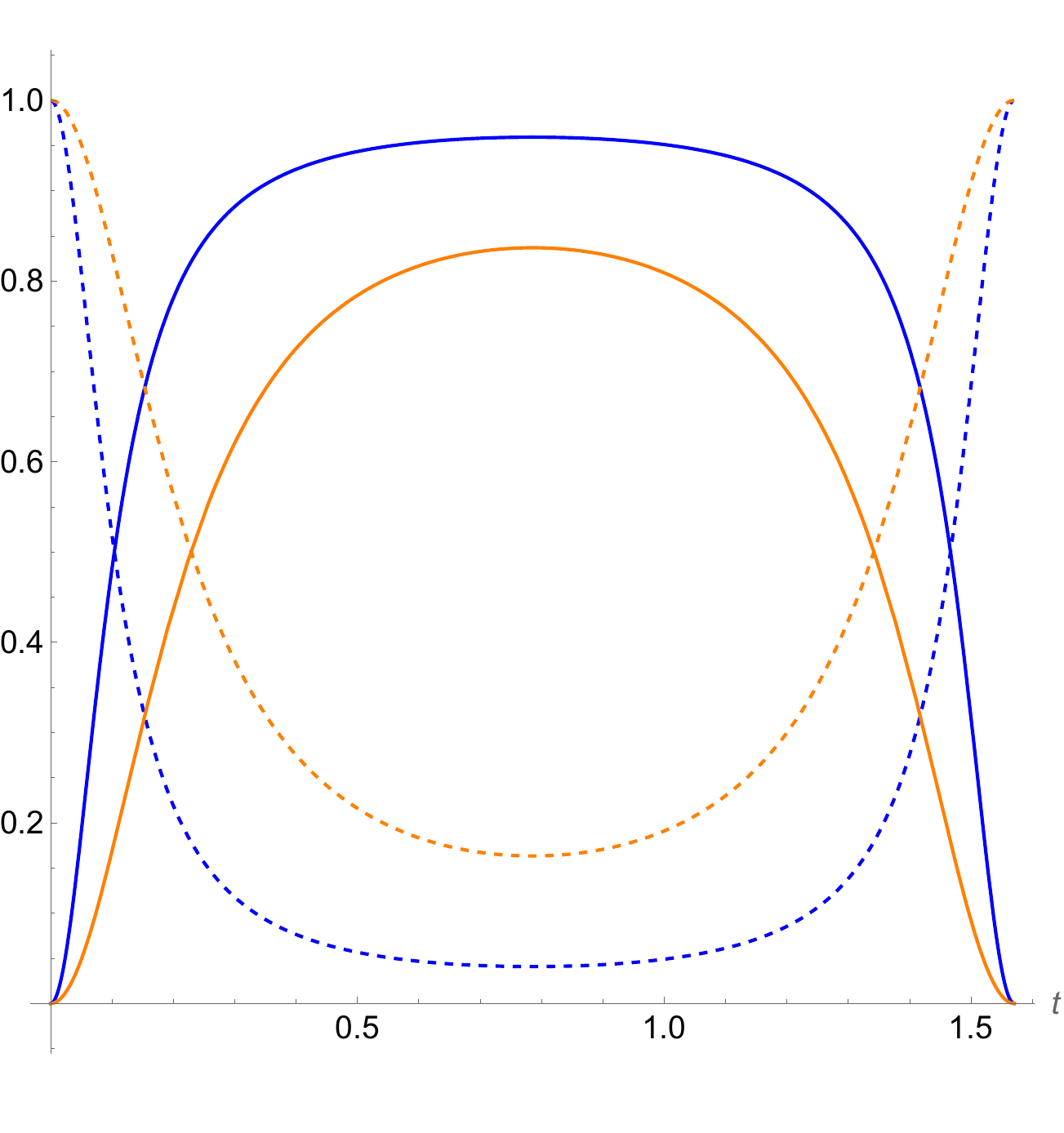}
    \includegraphics[width=7cm,height=7cm]{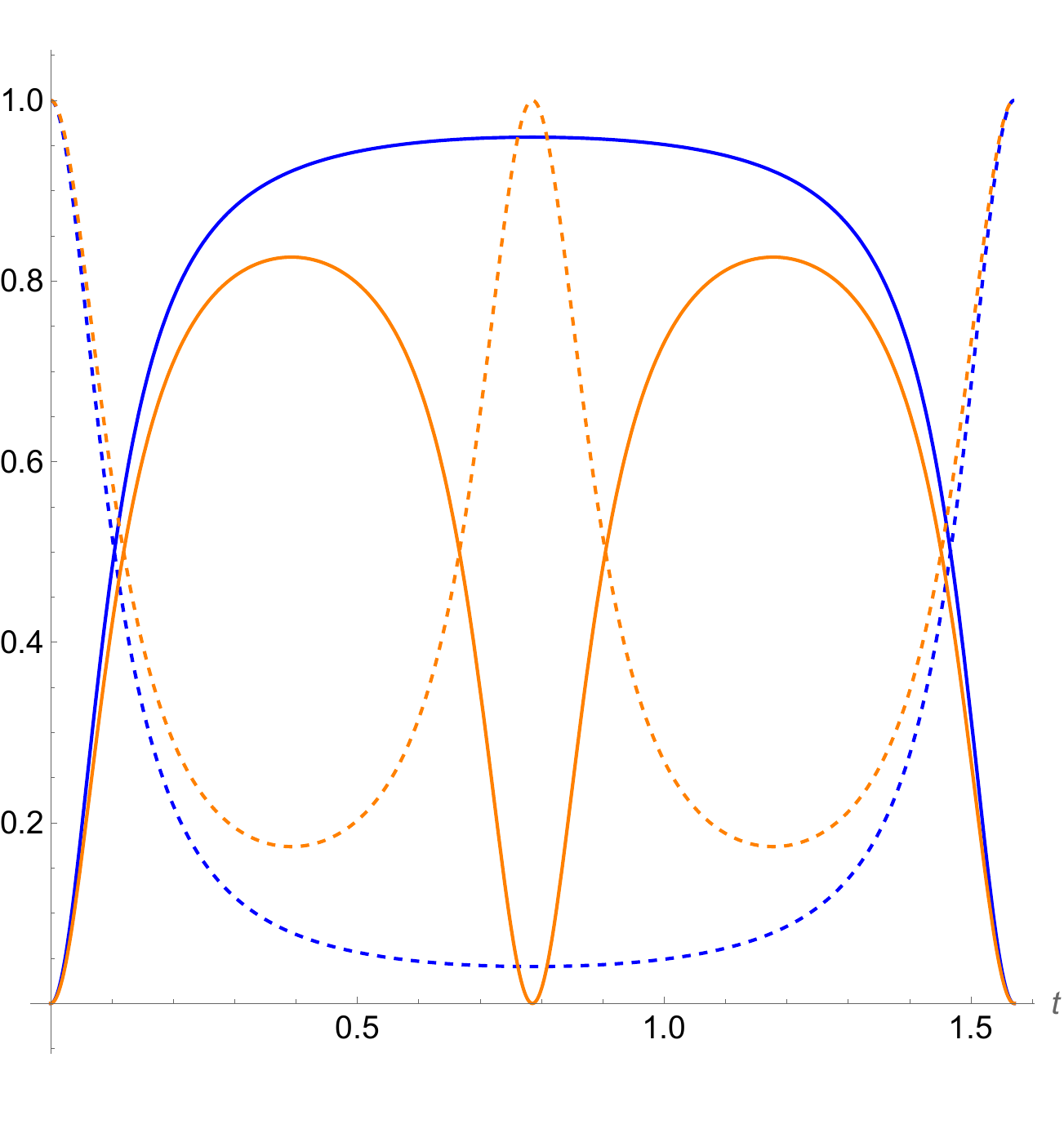}
     \caption{The left panel plots the radial $R_z$ (solid lines) and angular $R_{\phi}$ (dotted lines) 
     contributions for different values of $L_{\phi}$ (blue for $L_{\phi}=1$, 
     orange for $L_{\phi}=2$), with  $\mathcal{H}=10$ and $\ell=1$. 
     The right panel plots $R_z$ and $R_{\phi}$ for different values of $\ell$ (blue for $\ell=1$, 
     orange for $\ell=2$), with  $\mathcal{H}=10$ and $L_{\phi}=1$.}
   \label{Relative_PM_theta_pi/2}
\end{figure}

To evaluate the relative importance of each term in \eqref{prop_mom_v1}, in figure 
\ref{Relative_PM_theta_pi/2} we plot the ratios of the radial and angular contributions--
given by $R_{z} \equiv \frac{P_z}{P_{\bar \rho} } \, \frac{\partial {\dot z}}{ \partial {\dot {\bar \rho}}}$
and  $R_{\phi} \equiv  \frac{P_{\phi}}{P_{\bar \rho} }   \, \frac{\partial {\dot \phi}}{ \partial {\dot {\bar \rho}}}$, 
respectively--as functions of time. In the left panel the plots of the ratios are for fixed 
$\mathcal{H}$ and $\ell$ and we increase the value of $L_{\phi}$, 
while in the right panel we keep fixed $\mathcal{H}$ and  $L_{\phi}$ and we increase the value of $\ell$. 
The analysis reveals that the angular contribution is initially concentrated around the turning points of the 
radial trajectory  ($t \rightarrow  \eta \,  \pi/2 \ell$ for $\eta = 0,1,2, \cdots$), while the radial contribution 
dominates elsewhere. However, an increase in either the Coulomb scale $\ell$ 
or the angular momentum $L_{\phi}$ diminishes the importance of the radial component, while simultaneously enhancing the angular contribution.

The blue curve in figure  \ref{PM_CX_theta_pi/2}  corresponds to the case without rotation 
and is similar to the curve that is presented in 
\cite{Fatemiabhari:2025usn}. The expansions of the proper momentum at the endpoints of the particle motion
for the blue curve are
\begin{equation}
P_{\bar \rho} = - \sqrt{\mathcal{H}^2 -  \ell^2} 
\Bigg[ t - \frac{\ell^2}{6}\, t^3  + \cdots \Bigg] \quad \& \quad  
P_{\bar \rho} = \pm \sqrt{\frac{\mathcal{H}^2}{\ell^2} -1} 
\Bigg[ 1 - \frac{\ell^2}{2}\, \left(t-\frac{\pi}{2 \, \ell}\right)^2  +\cdots \Bigg] \, .
\end{equation}
The UV behavior, recovered in the limit $\ell \rightarrow 0$, coincides with results that were presented in \cite{Caputa:2024sux} and  \cite{Fatemiabhari:2025usn}. 
In contrast, the IR regime is significantly modified by the presence of the parameter $\ell$. 
Specifically, as $\ell \rightarrow 0$, the end of space recedes to \(z\rightarrow \infty \) and the momentum diverges, a feature that aligns with the original analysis in pure AdS ($\ell=0$) 
where the momentum exhibits a late-time divergence.

The presence of rotation modifies this behavior; specifically, the proper momentum approaches a constant value in the \(t\rightarrow 0\) limit. Regarding the IR regime, the behavior remains qualitatively similar to the non-rotating case, with the momentum diverging as \(\ell \rightarrow 0\). 
The expansions of the proper momentum at the endpoints of the particle's trajectory, for cases 
involving rotation, are given by
\begin{equation}
P_{\bar \rho} = - \sqrt{\mathcal{H}^2 \, z_-^2 -1}  + {\cal O} (t^2)
\quad \& \quad  
P_{\bar \rho} = \pm \sqrt{\mathcal{H}^2 \, z_+^2-  1}   + {\cal O} \left(t- \frac{\pi}{2 \, \ell}\right)^2 \, . 
\end{equation}

Complexity ${\cal C}(t)$ is proportional to the integral of the proper momentum
\begin{equation} \label{def_complexity}
{\cal C} = - \, \alpha \int P_{\bar \rho}  dt \quad \xRightarrow[]{\alpha=1} \quad 
{\cal C}= -  \int_{z_-} \frac{P_{\bar \rho}}{\dot z}  dz 
\end{equation}
and the constant of proportionality is set to unity. 
In \cite{Caputa:2024sux}, this constant is fixed to ensure consistency with the dual CFT results. 

In the non-rotating case, substituting the results from \eqref{prop_mom_v2} and  \eqref{z_dot_v2} 
into \eqref{def_complexity} allows the integral to be performed analytically, yielding
\begin{equation}
{\cal C} = \int_{\mathcal{H}^{-1}} \frac{\mathcal{H} \, z'}{\sqrt{1- \ell^2 \, z'^2}} dz' = - \, 
\frac{\mathcal{H}}{\ell^2} \Bigg[ \sqrt{1- \ell^2 \, z^2} -  \sqrt{1- \frac{\ell^2}{\mathcal{H}^2}} \Bigg]
\end{equation}
and substituting the analytic expression of $z$ as a function of $t$ from  \eqref{z_function_t_no_rotation}, 
we arrive to the following expression for the complexity in the non-rotating case
\begin{eqnarray} \label{complexity_no_rotation_theta_pi_over_2}
{\cal C}(t) &=& \frac{1}{\ell} \sqrt{\frac{\mathcal{H} ^2}{\ell^2}-1} \Big[1-\cos (t \ell) \Big] \quad {\rm for} 
\quad t \in \left[0, \frac{\pi}{2\, \ell}\right] 
\nonumber \\
{\cal C}(t) &=& \frac{1}{\ell} \sqrt{\frac{\mathcal{H} ^2}{\ell^2}-1} \Big[1+\cos (t \ell) \Big] \quad {\rm for} 
\quad t \in \left[\frac{\pi}{2\, \ell}, \frac{\pi}{\ell}\right] \, . 
\end{eqnarray}
In the right panel of figure \ref{PM_CX_theta_pi/2} the blue curve represents 
the complexity evolution in the absence of rotation.
Some comments are in order: The form of the plot is very similar to the plot of the complexity 
in the confining background of \cite{Anabalon:2021tua} that was presented in \cite{Fatemiabhari:2025usn}. 
The geometry terminates at a finite radial distance (the finiteness in \cite{Fatemiabhari:2025usn} is due to the confining scale $Q$ and in our case is due to the Coulomb scale $\ell$) and a particle that is released from rest near the boundary reaches the end of space within a finite time 
and bounces back to its starting point, leading to a periodic, oscillatory motion.
The proper momentum remains finite but undergoes a sign reversal upon reflection at the IR boundary and the  
complexity exhibits a similar oscillatory behavior. 

This phenomenon has been previously documented in the literature, see, for example, \cite{Caputa:2024sux,Baiguera:2025dkc} for CFTs on finite intervals, and such a 
behavior was attributed to the finite dimension of the underlying Hilbert space. 
Similar considerations apply to the present case.

While the dual theories to the confining background of \cite{Anabalon:2021tua} were shown to exhibit an IR gap \cite{Chatzis:2024top}, the present case involves a non-confining background. Nevertheless, a geodesic fixed at \(\theta =\pi /2\) experiences an effective confining potential (see \cite{Brandhuber:1999jr} 
for  the calculation of the Wilson-Maldacena loop that exhibits an area law), 
with a mass gap proportional to \(\ell \). Coupled with the initial condition near the boundary--which imposes a UV cutoff--this effectively restricts the state or operator to a finite-dimensional Hilbert space. Such finiteness naturally accounts for the oscillatory behavior observed in the complexity.

\begin{figure}[ht] 
   \centering
   \includegraphics[width=7cm,height=7cm]{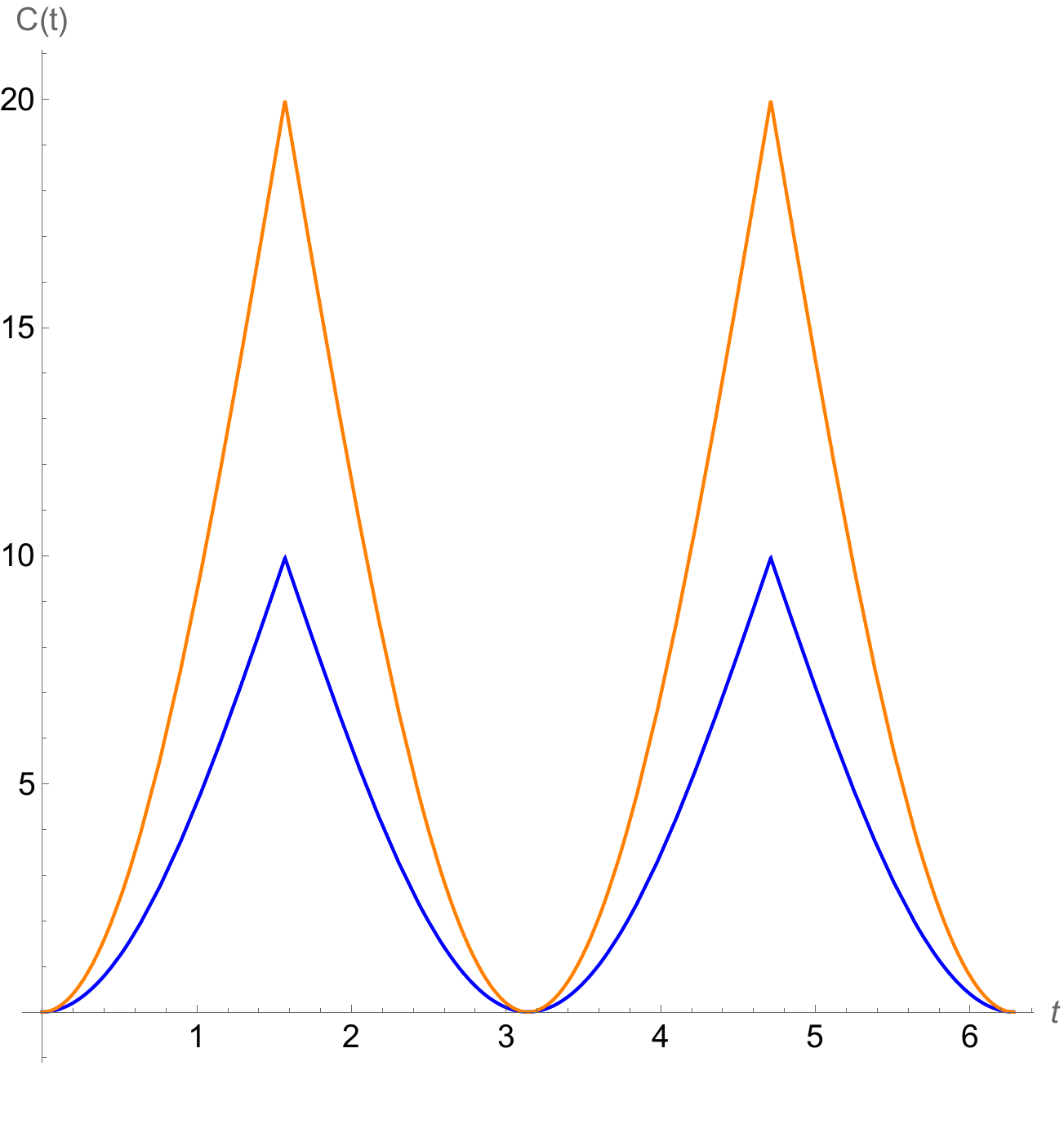}
    \includegraphics[width=7cm,height=7cm]{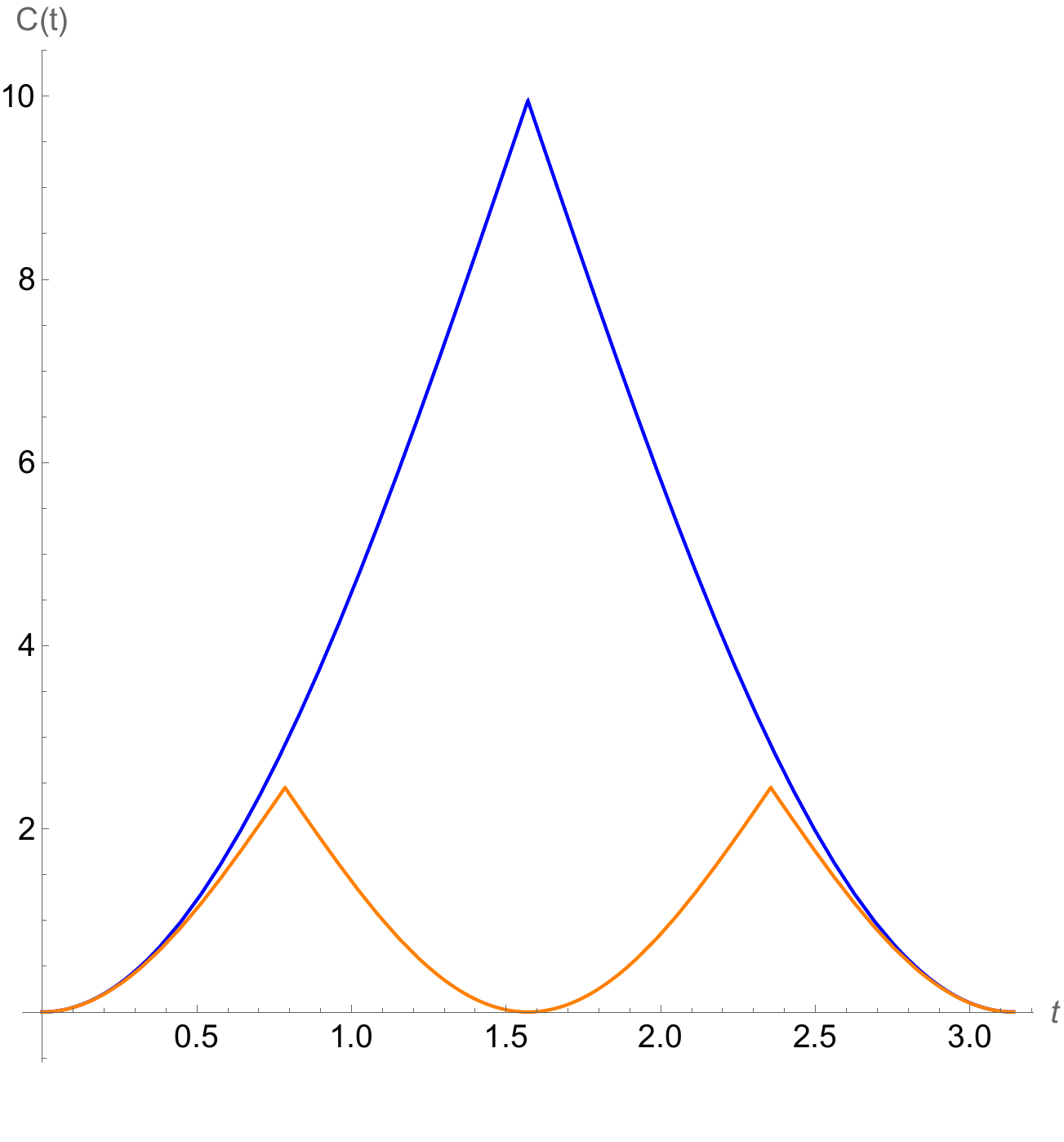}
     \caption{The left panel plots ${\cal C}(t)$ for $\ell=1$ and two different values of 
     $\mathcal{H}$  (blue for $\mathcal{H}=10$, orange for $\mathcal{H}=20$) 
     and the right panel ${\cal C}(t)$ for $\mathcal{H}=10$ and 
     two different values of $\ell$ (blue for $\ell=1$, orange for $\ell=2$).}
   \label{CX_no_rotation_theta_pi/2}
\end{figure}

In figure \ref{CX_no_rotation_theta_pi/2}, we further explore complexity evolution without internal rotation. 
The left panel shows that for a fixed  $\ell$, increasing the energy parameter $\mathcal{H}$
amplifies the maximum complexity.
In the right panel, where $\mathcal{H}$ is held constant, it becomes evident that a larger Coulomb parameter 
leads to suppressed amplitudes and increased oscillation frequencies.

In \cite{Fatemiabhari:2026goj} (see also \cite{Fatemiabhari:2025usn}) a comparison was put forward between
holographic complexity in a confining geometry and its counterpart in a modified (confining) version of the 
Ising model \cite{Jiang:2025wpj}. 
In the latter, the Krylov complexity--derived via the Lanczos algorithm--exhibits reduced amplitudes and increased 
frequencies as the confining parameter in the ferromagnetic phase grows. 
Given that the geodesic here experiences an effective confining potential, our observations in the right panel of 
figure \ref{CX_no_rotation_theta_pi/2} are qualitatively consistent with these results.

The analysis becomes even more interesting upon introducing rotation in the internal space. In the right panel 
of figure \ref{PM_CX_theta_pi/2} the red and the orange curves depict the complexity evolution 
for $L_{\phi}=2$ and $L_{\phi}=4$ respectively. As can be seen from the two plots of figure \ref{z_angle_theta_pi/2},
the coordinates \(z\) and \(\phi \) evolve in phase under rotation: both increase in the 
interval \(t\in [0,\pi /2\ell ]\) and subsequently decrease for \(t\in [\pi /2\ell ,\pi /\ell ]\). 
Consequently, one would expect the complexity to grow with increasing angular momentum, which is indeed observed.

Despite growing with angular momentum, the complexity always remains finite, in contrast to the divergence that is expected in pure AdS. 
Notice, that there exists an upper bound for \(L_{\phi }\) that depends on the values of \(\mathcal{H}\) and \(\ell \) 
(i.e.  $L_{\phi }< (\mathcal{H}^2 -\ell^2)/ 2  \mathcal{H}  \ell$); 
beyond this threshold, the proper momentum becomes complex, indicating a physical limit to the allowed rotation. 
This behavior is in agreement with the results that are reported in  \cite{Fatemiabhari:2026goj}. 
The complexity behavior that we observe, under the influence of rotation, is complementary to the findings of
\cite{Caputa:2024sux} and arrises from the use of a top-down construction within a well-defined Type IIB background. 
This is the proper framework to account for the influence of internal sphere rotation.

\subsection{Trajectory with $\theta=0$}
\label{subsec:PM_CX_theta_0}

For a radial trajectory at $\theta=0$, the geodesic distance in \eqref{10d_metric} simplifies to
\begin{equation} \label{geo_distance_theta_0}
ds^2 = \frac{\zeta}{z^2} \frac{dz^2}{\lambda^6}    +\frac{1}{\zeta} \, d\phi^2 \equiv  d{\bar \rho}^2 
\end{equation}

In the absence of rotation, it is possible to perform a similar analysis with the $\theta=\pi/2$ case, 
and the above relation becomes 
\begin{equation}
d{\bar \rho}^2 =   \frac{\zeta}{z^2} \frac{dz^2}{\lambda^6} \quad \Rightarrow \quad 
d{\bar \rho} = -  \frac{1}{z} \frac{dz}{\left(1 - \ell^2 \, z^2\right)^{1/4}} 
\quad \xRightarrow[]{z\rightarrow 0} \quad 
z \sim e^{-{\bar \rho }} \, . 
\end{equation}
Consequently, the limit ${\bar \rho} \rightarrow \infty$ maps to the UV boundary where 
$z \sim e^{-{\bar \rho }}$ holds, and the limit $\bar \rho \rightarrow 0$ corresponds to the IR scale $z_*=\frac{1}{\ell}$.

\begin{figure}[ht] 
   \centering
   \includegraphics[width=7cm,height=7cm]{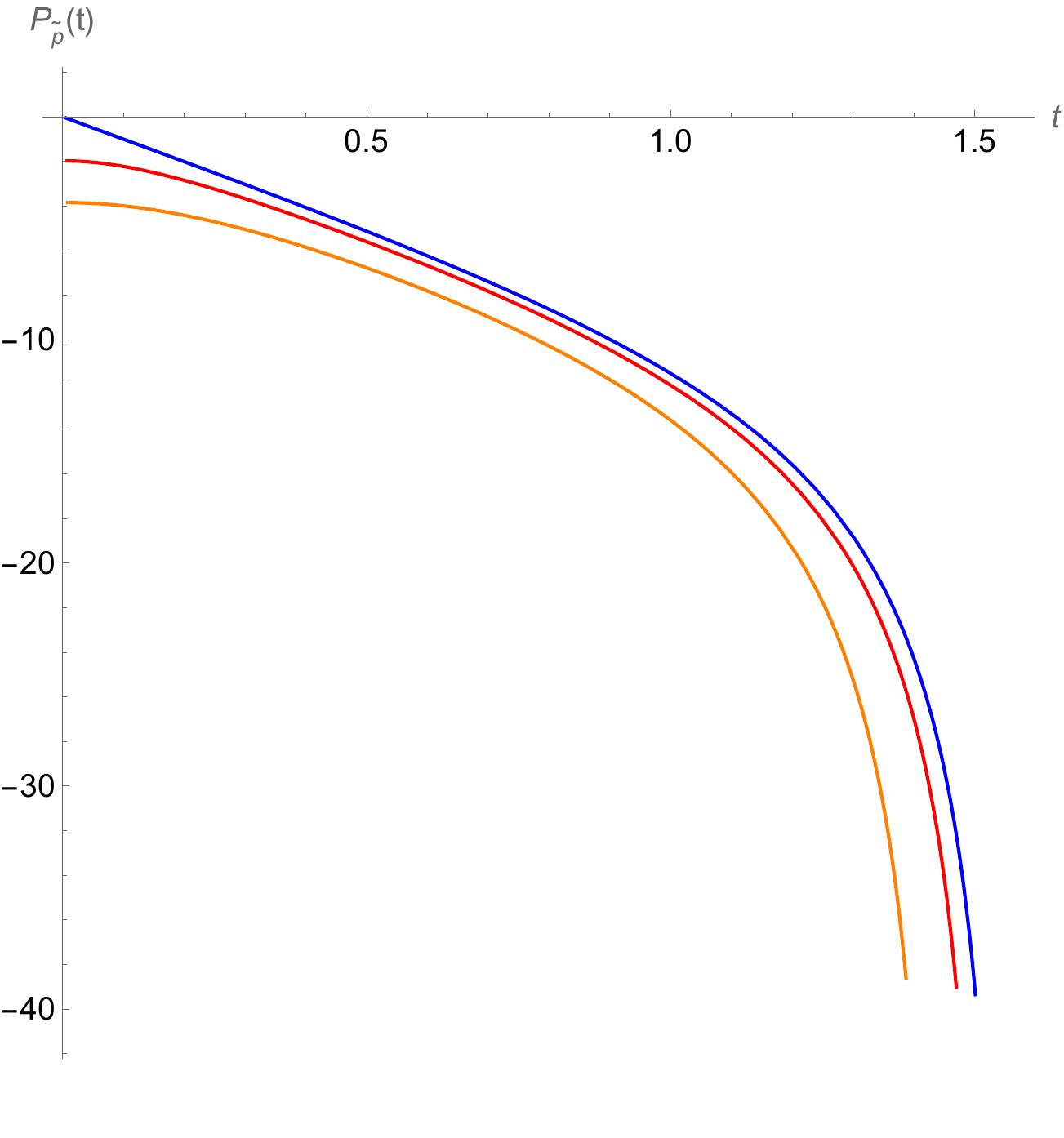}
   \includegraphics[width=7cm,height=7cm]{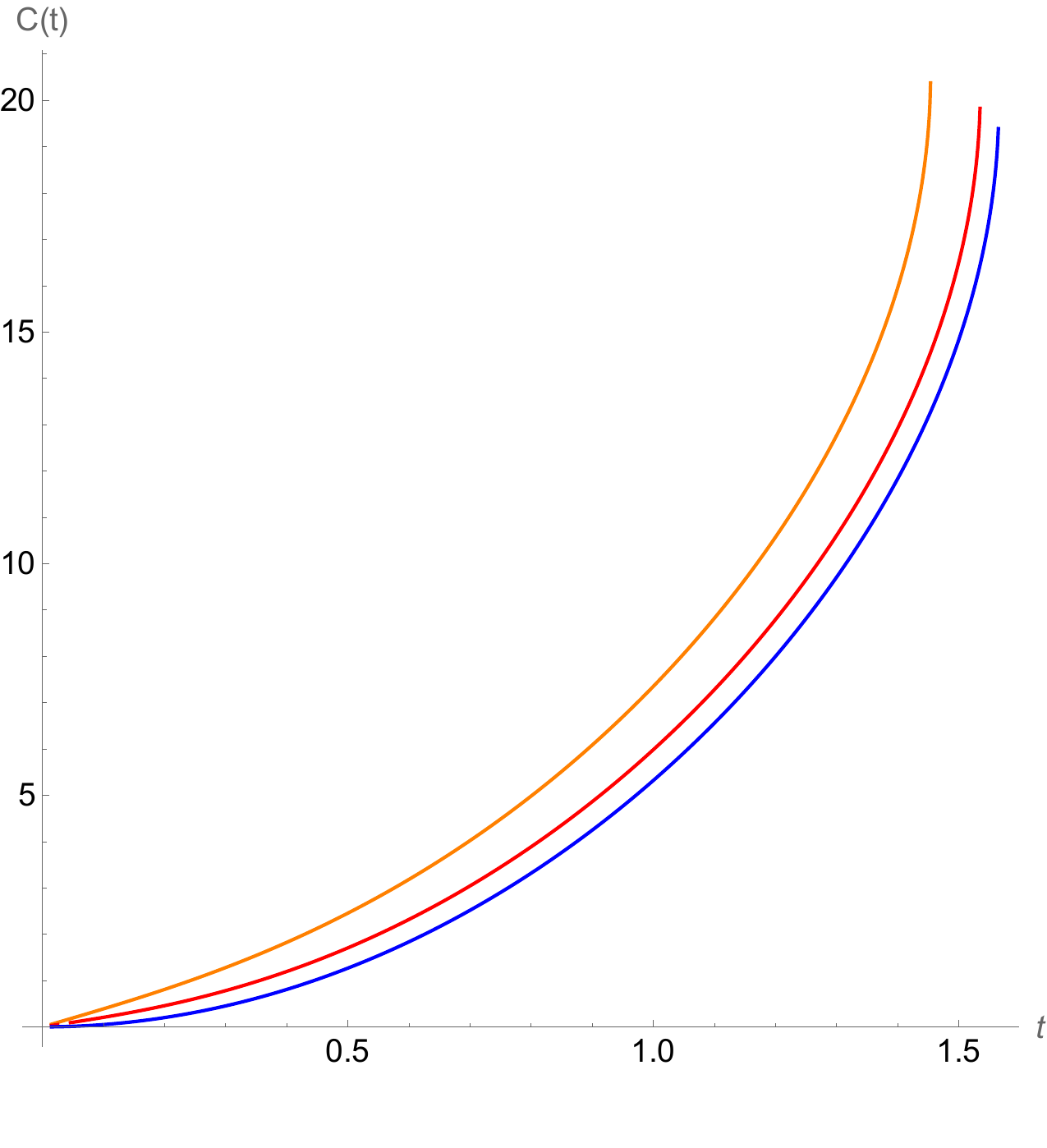}
     \caption{The left panel plots $P_{\bar \rho} $  and the right panel ${\cal C}(t)$
     for different values of $L_{\phi}$ (blue for $L_{\phi}=0$, red for $L_{\phi}=2$, 
     orange for $L_{\phi}=4$), with  $\mathcal{H}=10$ and $\ell=1$.}
   \label{PM_CX_theta_0}
\end{figure}

Following the analysis that is detailed in the previous subsection, we arrive to following expression 
for the proper momentum
\begin{equation} \label{prop_mom_theta_0_v2}
 P_{\bar \rho} =  - \frac{\sqrt{\mathcal{H}^2 \, z^2 -\sqrt{1 - \ell^2 \, z^2 }}}{(1 - \ell^2 \, z^2 )^{1/4}} \, . 
\end{equation}
Similarly to the $\theta=\pi/2$ case, the proper momentum depends only implicitly on the angular 
momentum, though the dependence of $z(t)$. An important difference with the $\theta=\pi/2$ case
is that the proper momentum diverges at $z=1/\ell$, and this divergence is there even if there is motion 
in the internal space. This is due to the fact that, with or without rotation, the end point of the radial 
trajectory is at $z=1/\ell$.
In the left panel of figure \ref{PM_CX_theta_0}
we have plotted the proper momentum as a function of time for different
values of the angular momentum $L_{\phi}$. 
The plot reveals two key behaviors: first, internal space motion enhances the proper momentum; 
second, the proper momentum diverges at the end of the radial trajectory. 
Increasing the angular momentum, the proper momentum diverges faster. 

\begin{figure}[ht] 
   \centering
   \includegraphics[width=7cm,height=7cm]{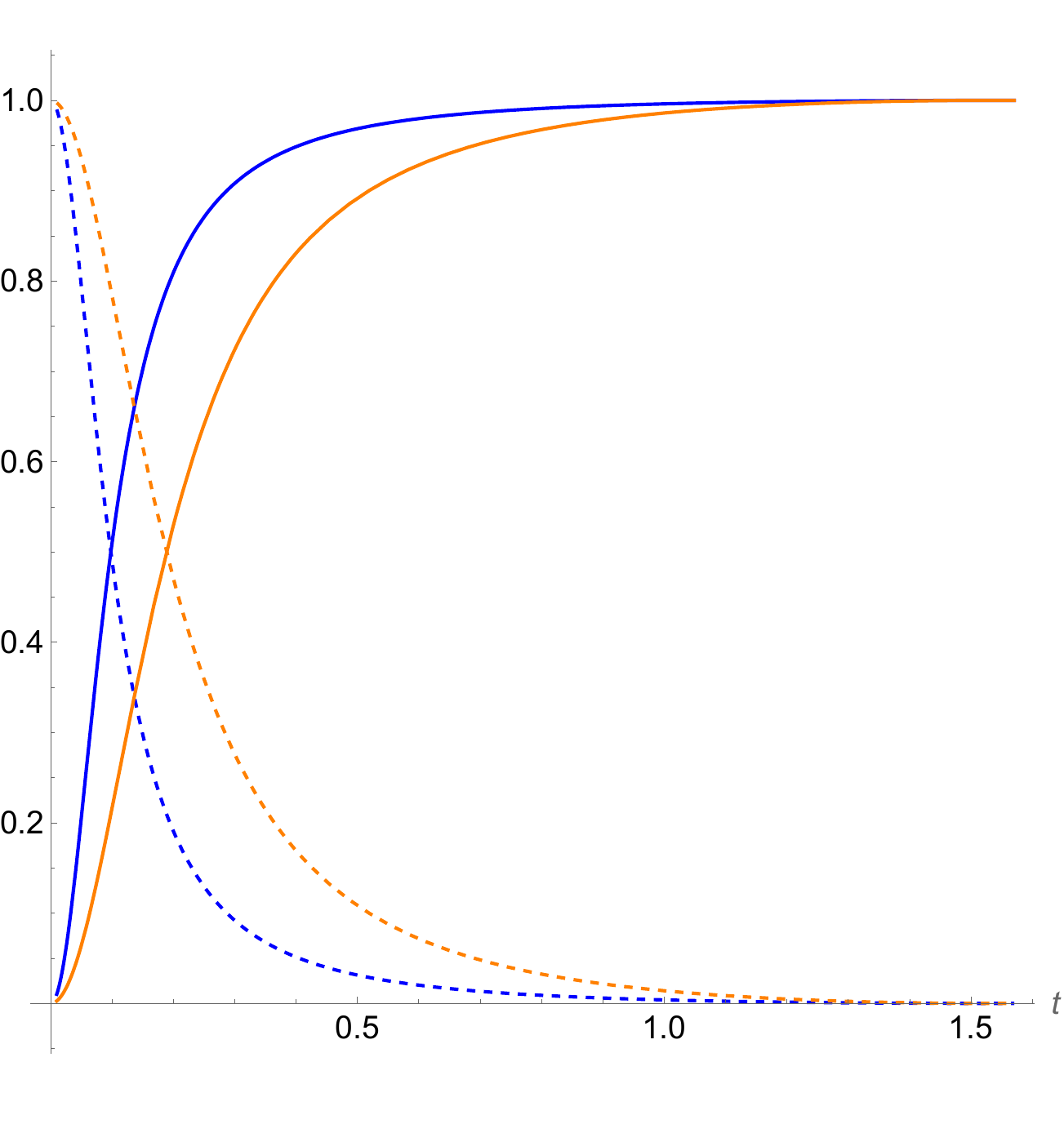}
    \includegraphics[width=7cm,height=7cm]{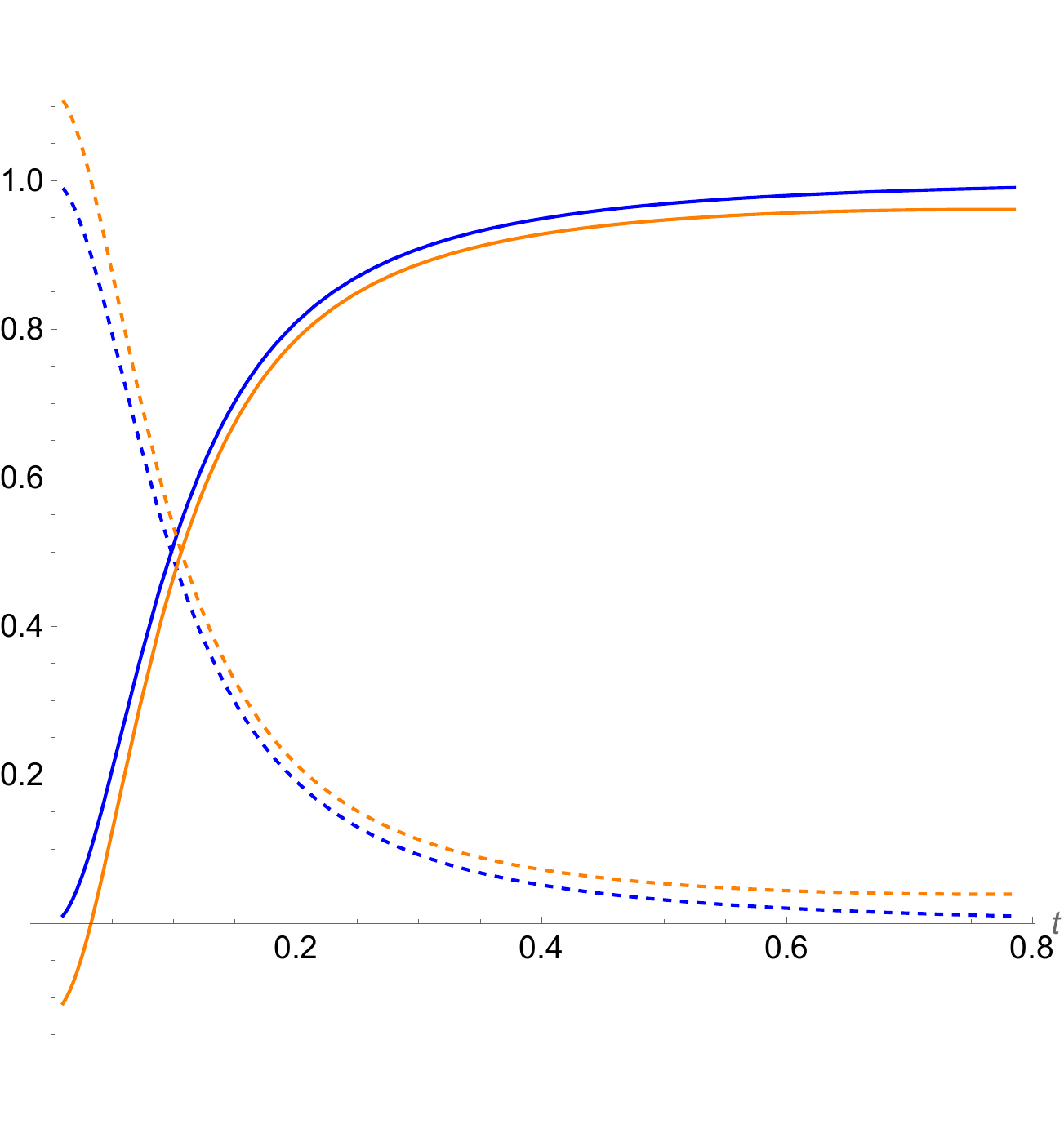}
     \caption{The left panel plots the radial $R_z$ (solid lines) and angular $R_{\phi}$ (dotted lines) 
     contributions for different values of $L_{\phi}$ (blue for $L_{\phi}=1$, 
     orange for $L_{\phi}=2$), with  $\mathcal{H}=10$ and $\ell=1$. 
     The right panel plots $R_z$ and $R_{\phi}$ for different values of $\ell$ (blue for $\ell=1$, 
     orange for $\ell=2$), with  $\mathcal{H}=10$ and $L_{\phi}=1$.}
   \label{Relative_PM_theta_0}
\end{figure}

In figure \ref{Relative_PM_theta_0}, following the same logic as the $\theta = \pi/2$ case, we 
illustrate the relative importance of each term, namely the radial and angular contributions, 
in the construction of  \eqref{prop_mom_theta_0_v2}. 
The left panel shows the ratios for fixed $\mathcal{H}$ and $\ell$
as $L_{\phi}$ increases, while the right panel depicts the effect of an increasing Coulomb scale 
$\ell$ for fixed $\mathcal{H}$ and  $L_{\phi}$. 
The analysis reveals that the angular contribution is dominating at early times while the radial contribution 
dominates at late times, i.e. for times approaching the maximum of the particle trajectory in the bulk. 
We have verified that this behavior persists if either the angular momentum or the Coulomb scale increase. 

In the case that there is no motion in the internal space, we can substitute the approximate expressions 
for $z(t)$ from \eqref{z_approx_no_rotation_theta_0} in \eqref{prop_mom_theta_0_v2} to obtain the 
following approximate expressions of the proper momentum, as functions of time, 
\begin{equation}
P_{\bar \rho} = \frac{2- \ell^2 \, (z^{0}_-)^2}{2 \, z^{0}_- \, \sqrt{1- \ell^2 \, (z^{0}_-)^2}} \, t + \cdots
\quad \& \quad 
P_{\bar \rho} = \pm \frac{\mathcal{H}}{\ell^{3/4}}\frac{1}{\sqrt{t- t_e}} + \cdots
\end{equation}
where the expressions for  $t_e$ and $z^{0}_-$ are given in \eqref{z_approx_no_rotation_theta_0}  and 
\eqref{z0initial_no_rotation_theta_0}, respectively. 
From those expressions it is evident that proper momentum is zero at the initial point of the radial trajectory, 
while it diverges when it approaches the ``terminal point", $z=1/\ell$. 

In the non-rotating case, substituting the results from \eqref{prop_mom_theta_0_v2} and  \eqref{z_phi_theta_0_dot_v1} 
into \eqref{def_complexity} allows the integral to be performed analytically, yielding
\begin{equation} \label{complexity_no_rotation_theta_0}
{\cal C} = \frac{2 \, \mathcal{H}}{\ell^2} \left[1- \ell^2 (z^{0}_-)^2 \right]^{1/4} - \frac{2 \, \mathcal{H}}{\ell^2} \left[1- \ell^2 \, z^2 \right]^{1/4}  
\end{equation}
and substituting the expression of $z$ as a function of $t$ from solving numerically  \eqref{integral_z_theta_0}, it is possible to 
obtain complexity as a function of time. From this expression we can see that complexity vanishes at $t=0$, while it is finite at $t=t_e$ 
(the second term in \eqref{complexity_no_rotation_theta_0} at  $t=t_e$ vanishes, since $z(t_e)=1/\ell$).
Even if the proper momentum diverges at  $t=t_e$, complexity is finite. 
In the right panel of figure \ref{PM_CX_theta_0} the blue curve represents 
the complexity evolution in the absence of rotation, while the red and the orange curves depict the complexity evolution 
for $L_{\phi}=2$ and $L_{\phi}=4$ respectively.

The presence of the curvature singularity at $z=1/\ell$, when $\theta=0$, is reflected in the divergence of the proper momentum. 
It should be emphasized that although the complexity remains finite, the results are not reliable in the vicinity of the singularity; 
our analysis should be restricted to the region far from it.
We have verified that this feature persists even for rotations within AdS; 
provided that $\theta=0$, the turning point of the geodesic is always at $z=1/\ell$. 
To resolve this issue, a new scale should be introduced. 
By combining the Coulomb with the confining deformation of \cite{Anabalon:2021tua} 
(using the background constructed in \cite{Anabalon:2024che}), it is plausible that
the geodesic will bounce off the confining scale before reaching the singularity. 
 
The results we are presenting are in agreement with those of 
\cite{Fatemiabhari:2025usn,Fatemiabhari:2026goj}. Specifically, in the $\theta =\pi/2$ case--where the IR is smooth-- 
the complexity exhibits an oscillatory behavior. Conversely, for $\theta =0$,
the presence of the curvature singularity in the IR prevents the emergence of complexity oscillations.

\subsection{Comparison with field theory}
\label{subsec:FTcomparison}

According to \cite{Nandy:2024evd,Anegawa:2024wov}, in order to reproduce the holographic 
result in  \eqref{complexity_no_rotation_theta_pi_over_2} through a field theory calculation, 
one must first determine the glueball mass spectrum.
In \cite{Anegawa:2024wov}, to describe a theory with a discrete spectrum and a mass gap,  
the authors consider a model of many free scalar fields in compact space, 
with Lagrangians of the following form
\begin{equation}
L \sim \sum_{n=0}^{\infty} \partial^{\mu} \phi_n \, \partial_{\mu} \phi_n - m_n^2 \,  \phi_n^2 \, . 
\end{equation}
The Krylov complexity is evaluated for the composite operator ${\cal O} = \sum \phi_n$.
In the context of QCD, this is illustrated considering a color-singlet operator, such as 
${\cal O} = {\rm Tr}[F^{\mu \nu} F_{\mu \nu}]$, where the components 
$\phi_n$ represent glueball states of varying masses.
In the large-$N$ limit these are treated as free fields, and both the Krylov complexity and the 
Lanczos coefficients follow directly from a given two-point function. 
For a large number of states the Krylov complexity has an oscillating behavior but 
the calculation lacks a closed-form analytic expression. 
However in the case of one glueball the computation is analytic and  
the Krylov complexity is proportional to $ \sin^2 (M t)$, where $M$ is the 
glueball mass. While the gravity background includes various glueball spins, a representative subsector could provide a sufficient estimate.

Following this rationale and employing the results from \cite{Brandhuber:1999jr}, 
we conjecture that the Krylov complexity is dominated by the lightest glueball state  with mass $M=\ell$. 
By identifying this mass with the Coulomb scale, 
the contribution of the scalar mode to the complexity is expected to be proportional to
 $ \sin^2 (\ell \, t)$, exhibiting an oscillating behavior with a period of $\pi/\ell$. 
The calculations from the gravity side for the $\theta =\pi/2$ case show a period of $\pi/\ell$, 
leading to a match between the gravity and field theory results.

In conclusion, we observe a qualitative agreement in the oscillation frequency of the complexity between the field theory and gravity computations, achieved by considering only the scalar glueball mass. It is worth noting that since the analysis in \cite{Anegawa:2024wov} is conducted on a compact manifold, a consistent comparison requires a gravity dual where one direction is compactified. Such a setup was explored in \cite{Fatemiabhari:2025usn} for a confining, gapped theory with an angle direction compactified on a circle. In that context, the confinement scale is related to the periodicity of the compact angle, and the authors find a qualitative agreement with the results of \cite{Anegawa:2024wov}, as both frameworks exhibit an oscillating behavior for the complexity.

 In the present setup, we don't have either a confining geometry or a compact direction. However, since the geodesic experiences an effective confining potential, we are applying the expression of \cite{Anegawa:2024wov} for the mass of the lightest glueball, and we find a match in the frequencies. 
This suggests that the complexity of the operator is sensitive to the discrete nature of the spectrum (mass gap) rather than the global topology of the manifold. As long as the operator excites localized, bound states, the Hilbert space dynamics effectively mimic those of a compact space.

\section{Conclusions and future directions}
\label{sec:conclusions}

The Coulomb branch of ${\cal N}=4$ SYM is the perfect arena to test and extend the recent proposal that relates the time derivative of the Krylov complexity 
with the proper momentum of a massive bulk particle. From one side there is the presence of UV cutoff and an IR end of space ensuring a finite Hilbert space truncation,
with whatever this may entail, and from another side there is curvature singularity at the end of space. This motivates the analysis of the Krylov complexity in a fully top-down framework. 

To fully exploit the plethora of scenarios that the Coulomb branch is offering for the study of complexity, we analyze the radial geodesics for two different radial trajectories. 
In the first one, the geodesic is sitting at $\theta=\pi/2$ and with this orientation the particle 
is not ``feeling" the curvature singularity, that is due to the multicenter distribution of branes. 
We have calculated analytically the particle trajectory and while the UV behavior reproduces 
the standard AdS result, the IR structure is drastically modified by the presence of the 
finite endpoint, which is characterized by the Coulomb scale. 
In the $\theta=0$ case similar calculations have been performed and the radial trajectory of the particle 
is qualitatively similar to the $\theta=\pi/2$ case. 
It exhibits a sinusoidal motion between a minimum and a maximum, and the fact that it is approaches 
the singularity in the far IR does not seem to produce an unusual feature.  
In both cases we have extended the analysis to geodesics with a non-trivial angular momentum 
dependence in the internal space.

Having obtained the geodesic equation for the particle trajectory, we move to the next step, 
which is the computation of the proper radial momentum and the Krylov complexity. 
Here our findings show a clear distinction between the two cases/trajectories: In the $\theta=\pi/2$ case,
the proper momentum remains finite and the Krylov complexity exhibits an oscillatory behavior. 
Within the holographic framework, these oscillations have a 
geometric origin. The motion of the massive probe is bounded between a UV cutoff and a 
IR end of space, that is characterized by the Coulomb scale. 
The amplitude and the frequency of those oscillations depend on the UV cutoff, the Coulomb scale and 
on the angular momentum arising from motion in the internal space.
Generic features are the following: 
Increasing the Coulomb scale, is decreasing the amplitude and is increasing the frequency of the oscillations. 
Moreover, increasing the value of the angular momentum, is increasing the amplitude complexity.

In the $\theta=0$ case, the proper momentum diverges, as the particle trajectory approaches the 
curvature singularity at $z=1/\ell$.  Consequently, even if the complexity remains finite, 
the results are unreliable in the vicinity of the singularity and there is no oscillatory pattern. 
Therefore, the analysis is limited to regions well away from the singularity.

The difference in the behavior of the geodesics between the two trajectories is also reflected in 
the temporal evolution of the radial and angular contributions to the proper momentum, 
as depicted in figures 
\ref{Relative_PM_theta_pi/2} and \ref{Relative_PM_theta_0}. When complexity exhibits an 
oscillatory pattern and the proper momentum is finite, the angular contribution dominates  
around the turning points of the radial trajectory, while the radial contribution dominates elsewhere. 
Increasing, either the Coulomb scale $\ell$ 
or the angular momentum $L_{\phi}$ diminishes the importance of the radial component, 
even at intermediate time scales. 
In the absence of complexity oscillations and with a divergent proper momentum, 
the angular contribution is dominating at early times while the radial contribution takes over at late times, 
i.e. for times approaching the singularity. This picture remain unchanged 
by an increase in either the angular momentum or the Coulomb scale.

Finally we have attempted a qualitative comparison between the field theory 
calculations of complexity and their holographic counterparts. 
On the field theory side, if one assumes that only a scalar glueball contributes to the Krylov complexity, 
then it is proportional to  $\sin^2 (\ell \, t)$, where $\ell$ represents the glueball mass.
We have shown that for $\theta = \pi/2$, holographic complexity reproduces not only 
the oscillating behavior but also the frequency of the field theory result. 
This provides a qualitative agreement between the two frameworks; 
it would be interesting to extend this comparison to the amplitude of the oscillations.


From the gravity side, it would be interesting to combine the Coulomb with the 
confining deformation of \cite{Anabalon:2021tua} (see the background in \cite{Anabalon:2024che}), 
to explore the behavior of the complexity. In this framework, the confining scale could ``hide" the 
curvature singularity at $z = 1/\ell$ for a geodesic that is sitting at $\theta=0$. 
Moreover, it would be interesting to calculate the complexity in 
other deformed ${\cal N}=4$ backgrounds, such as the marginally deformed ${\cal N}=4$ \cite{Lunin:2005jy} 
(see also \cite{Hernandez:2005xd,Hernandez:2005zx} for the Coulomb branch of the marginally deformed and \cite{Hammond:2026lle} for a combination of confining/Coulomb/marginally deformed background) and see the effects of the deformation in the behavior of the complexity.

In the spirit of investigating competition of scales in the behavior of complexity, it would be natural to examine the effect of a second angular momentum. In the calculation that we have presented, the angular momentum is due to a motion inside the $S^5$ and one could also add a rotation inside the AdS part. Notice here that in 
\cite{Fatemiabhari:2025usn}, with which our results are in qualitative agreement, the angular momentum 
is for a geodesic motion inside the AdS. However, a competition of scales 
(between the two angular momenta) may emerge. 


\subsection*{Acknowledgements}

We are grateful to Dimitrios Chatzis, George Georgiou, Carlos Nunez and 
Konstantinos Sfetsos for carefully 
reading the manuscript and for providing insightful comments.
This paper has been financed by the funding programme ``MEDICUS", of the University of Patras (grant number: 83800).


\bibliographystyle{utphys}

\bibliography{refs}

\end{document}